\documentclass[jcp, twocolumn, superscriptaddress]{revtex4-1}
\newcommand{\SingleColFigScale}{0.9} 

\usepackage{graphicx}
\usepackage[usenames,dvipsnames]{xcolor}
\usepackage{amsmath}
\usepackage{amssymb}
\usepackage{mathtools}
\usepackage{accents}
\usepackage{bm}
\usepackage[pdfusetitle]{hyperref}
\definecolor{link}{rgb}{0.07, 0.07, 0.80}
\hypersetup{
  colorlinks=true,
  linkcolor=link,
  citecolor=link,
  urlcolor=link
}

\makeatletter
\def\blfootnote{\gdef\@thefnmark{}\@footnotetext}
\makeatother

\begin{document}

\title{Generalization of the hierarchical equations of motion theory for efficient calculations with arbitrary correlation functions}

\blfootnote{Corresponding authors: T.~Ikeda and G.~D.~Scholes}

\author{Tatsushi Ikeda}
\email{tikeda@princeton.edu, t\_ikeda@chemsys.t.u-tokyo.ac.jp}
\thanks{Present address: Department of Chemical System Engineering, The University of Tokyo, Tokyo 113-8656, Japan }
\affiliation{Department of Chemistry, Princeton University, Washington Road, Princeton, New Jersey, 08544, USA}
\author{Gregory D. Scholes}
\email{gscholes@princeton.edu}
\affiliation{Department of Chemistry, Princeton University, Washington Road, Princeton, New Jersey, 08544, USA}

\date{\today}

\begin{abstract}
  The hierarchical equations of motion (HEOM) theory is one of the standard methods to rigorously describe open quantum dynamics coupled to harmonic environments.
  Such a model is used to capture non--Markovian and non--perturbative effects of environments appearing in ultra--fast phenomena.
  In the regular framework of the HEOM theory, the environment correlation functions are restricted into linear combinations of exponential functions.
  In this article, we present a new formulation of the HEOM theory including treatments of non--exponential correlation functions, which enables us to describe general environmental effects more efficiently and stably than the original theory and other generalizations.
  The library and its Python binding we developed to perform simulations based on our approach, named LibHEOM and PyHEOM respectively, are provided as supplementary material.
  
  \noindent\hrulefill
  
  This article may be downloaded for personal use only.
  Any other use requires prior permission of the author and AIP Publishing.
  This article appeared in [T.~Ikeda and G.~D.~Scholes, J.~Chem.~Phys.~\textbf{152}, 204101 (2020)] and may be found at \href{https://doi.org/10.1063/5.0007327}{https://doi.org/10.1063/5.0007327}.
\end{abstract}

\pacs{Valid PACS appear here}


\maketitle

\section{INTRODUCTION}
\label{sec:introduction}
Open quantum theories, i.e., theoretical descriptions for quantum states/dynamics of a system exposed to fluctuation and dissipation of surrounding environments, are important subjects in a wide range of physics and chemistry because the environmental effects are ubiquitous.
Recent advances in experimental technologies have made it possible to observe electronic/vibrational dynamics in very short time and atomic spatial scales in which quantum properties of materials (e.g., quantum superposition and quantum tunneling) plays essential roles \cite{scholes2017nature, scholes2018jpcl, mancal2020cp}, and quantum theories becomes more significant as theoretical inputs to analyze such experiments \cite{ishizaki2007jpca, kreisbeck2012jpcl, dean2016chem, prokhorenko2016jpcl, miyata2017mc, rafiq2019jacs}.
In terms of dynamics, various quantum/semi--classical theories have been constructed to capture the quantum nature, including equations of motion for wave functions, density matrices, phase space distributions \cite{redfield1965inbook, tanimura1989jpsj, tanimura1994jcp, zhang1998jcp, kapral1999jcp, thoss2000jcp, kuhl2002jcp, xie2014jcp, ikeda2017jcp, ikeda2019jctc, chen2019jpcl}, and Gaussian quantum wavepackets \cite{ben1998jcp, ben2000jcpa, makhov2014jcp}, and approaches utilizing mixed quantum--classical trajectories \cite{tully1990jcp, hammes1994jcp, coker1995jcp, stock2005acp, subotnik2016arpc}.

Open quantum theories are needed to predict and analyze spectra of advanced non--linear spectroscopy experiments.
These spectra occasionally show complicated features caused by transitions and beatings among many quantum states and non--Markovian and non--perturbative effects of environments appear as, e.g., time--dependent Stokes shift and spectral diffusion, which requires more rigorous theories for prediction and interpretation \cite{tanimura2009acr, ishizaki2008cp, fujihashi2015jcp}.
Open quantum theories offering rigorous non--Markovian/non--perturbative numerical results include the quasiadiabatic propagator path integral (QUAPI) \cite{makri1992cpl, ilk1994jcp}, mapping Hamiltonian approach with the density matrix renormalization group (DMRG) \cite{chin2010jmp, prior2010prl}, multi--configuration time--dependent Hartree (MCTDH) theory \cite{meyer1990cpl, manthe1992jcp, wang2000jcp, wang2007jpca}, and hierarchical equations of motion (HEOM) theory \cite{tanimura1989jpsj, ishizaki2009jcp}.
In this article, we discuss a new generalization of the HEOM theory.

The HEOM theory was originally developed by Tanimura and Kubo to describe the dynamics of an open quantum system coupled to a high--temperature Drudian environment which could be characterized by a single--exponential environment correlation function \cite{tanimura1989jpsj}.
This theory was later extended into the cases for multi--exponential correlation functions \cite{tanimura1990pra, tanimura1994jpsj, ishizaki2005jpsj, tanaka2009jpsj, tanimura2012jcp}.
Hereafter, we refer to an element of the correlation function decomposition as a basis.
Although this original HEOM theory enables us to describe the effects of non--perturbative and non--Markovian system--environment coupling for environment correlation given in the form of a multi--exponential basis set, it may still fail under circumstance when an exponential function basis set is not the optimal expression of the correlation function, or when it is impossible to decompose the correlation functions into the basis set.

A generalization of the HEOM theory for a non--exponential basis set is given in an earlier work by Yan and co--workers \cite{xu2005jcp}.
Another generalization is made by Wu and co--workers in a systematic way as the extended HEOM \cite{tang2015jcp, duan2017prb}.
However, these generalizations have redundancies and are numerically challenging:
These generalizations lead up to a $2K$ dimensional hierarchy for the case of a $K$ general function basis, whereas the original HEOM theory forms a $K$ dimensional hierarchy in its time--dependent differential equations in the case of a $K$ exponential function basis set.
Such an increase in the hierarchy dimension results in huge demands of computational memory and long computational time.
Moreover, in the above generalization, the hierarchy structures for an exponential function basis are different from those for a non--exponential function basis, which make it difficult for discussing them in a unified framework.
Although a new approach including treatments using a non--exponential basis set with $K$ dimensional hierarchy has been recently proposed, the hierarchical Schr\"{o}dinger equations of motion (HSEOM) approach \cite{nakamura2018pra}, this theory has a limitation in the possible forms of basis functions, and is unstable in its long--time behavior.

In this article, we present a new generalization of the HEOM theory including treatments of non--exponential correlation functions, which could be more efficient and stable than the original theory and other generalizations.
Our generalization has a similar hierarchy structure to the original HEOM theory, and therefore it is easy to implement the theory by using a similar code to the original theory.
We demonstrate our new generalization by using three examples that demonstrate non--exponential behavior of the environment correlation functions.

The organization of this paper is as follows.
In section \ref{sec:theory}, we show our generalization and its relation to other theories.
In section \ref{sec:results}, we present numerical results for cases of super--Ohmic spectral density, critically--damped Brownian spectral density, and zero--temperature environment as demonstrations.
Section \ref{sec:conclusion} is devoted to concluding remarks.
The library and its Python binding we developed to perform simulations based on our approach, named LibHEOM and PyHEOM respectively, are provided as supplementary material, of which up--to--date versions may be found on GitHub.

\section{THEORY}
\label{sec:theory}

In this section, we show our generalization of the HEOM theory.
To construct the original HEOM theory, the path integral formulation with the Feynman--Vernon influence functional is typically used \cite{tanimura1989jpsj, tanimura2006jpsj}.
Here, we employ a cumulant expansion technique, which is used in Ishizaki and Fleming's work \cite{ishizaki2009jcp}.

\subsection{Hamiltonian}

We consider a system linearly coupled to a harmonic environment (bath).
Here, we assume that the system--bath interaction is characterized by a single system subspace operator $\hat{V}$ for simplicity.
An extension for multiple operators with multiple independent environments is trivial, and has been implemented in our codes (See supplementary material).
The total Hamiltonian of the system is expressed as
\begin{align}
  \hat{H}^{\mathrm{tot}}&\equiv \hat{H}+\hat{H}^{\mathrm{bath}}+\hat{H}^{\mathrm{int}}
  \label{eq:total-hamiltonian},
\end{align}
where $\hat{H}$, $\hat{H}^{\mathrm{bath}}$, and $\hat{H}^{\mathrm{int}}$ are the Hamiltonians of the system, bath, and interaction, respectively.
\begin{subequations}
  The bath Hamiltonian reads
  \begin{align}
    \hat{H}^{\mathrm{bath}}\equiv \sum _{\xi }\frac{\hbar \omega _{\xi }}{2}\left(\hat{p}_{\xi }^{2}+\hat{x}_{\xi }^{2}\right),
  \end{align}
  where $\hat{x}_{\xi }$, $\hat{p}_{\xi }$, and $\omega _{\xi }$ are the dimensionless coordinate, conjugate momentum, and characteristic frequency of the $\xi $th bath mode, and the interaction Hamiltonian is expressed as
  \begin{align}
    \hat{H}^{\mathrm{int}}\equiv -\sum _{\xi }g_{\xi }\hat{x}_{\xi }\hat{V},
  \end{align}
  where $g_{\xi }$ is the coupling strength between the system and $\xi $th bath model.
\end{subequations}
When our system is a spin system, the Hamiltonian $\hat{H}^{\mathrm{tot}}$ corresponds to the well--known spin--boson model \cite{leggett1987rmp}.

\subsection{Cumulant expansion of time evolution equation}
To evaluate the effects of the interaction Hamiltonian $\hat{H}^{\mathrm{int}}$ to the dynamics, we introduce the interaction picture based on the non--interacting Hamiltonian, $\hat{H}^{0}\equiv \hat{H}+\hat{H}^{\mathrm{bath}}$, as
\begin{align}
  \tilde{O}(t)\equiv e^{+i\hat{H}^{0}(t-t_{0})/\hbar }\hat{O}(t)e^{-i\hat{H}^{0}(t-t_{0})/\hbar }.
\end{align}
Here, a tilde on an operator indicates the operator is in the interaction picture.
In this picture, the time evolution equation of the total density operator, i.e., the Liouville--von Neumann equation, can be written as $\partial _{t}\tilde{\rho }^{\mathrm{tot}}(t)=-\tilde{\mathcal{L}}^{\mathrm{int}}(t)\tilde{\rho }^{\mathrm{tot}}(t)$, and the solution is
\begin{align}
  \tilde{\rho }^{\mathrm{tot}}(t)=\mathcal{T}_{+}\exp \left(-\int _{t_{0}}^{t}\!ds\,\tilde{\mathcal{L}}^{\mathrm{int}}(s)\right)\tilde{\rho }^{\mathrm{tot}}(t_{0}).
\end{align}
Here, we have defined a Liouvillian of a Hamiltonian $\hat{H}$ by $\mathcal{L}\equiv i[\hat{H},\dots ]/\hbar $, and have introduced the chronological time ordering operator $\mathcal{T}_{+}$.
We assume that the total density operator at $t=t_{0}$ can be written as
\begin{align}
  \hat{\rho }^{\mathrm{tot}}(t_{0})=\hat{\rho }(t_{0})\otimes \hat{\rho }_{\mathrm{eq}}^{\mathrm{bath}},
  \label{eq:fact-init}
\end{align}
where $\hat{\rho }(t_{0})=\tilde{\rho }(t_{0})$ is reduced density operator of the system subspace and $\hat{\rho }_{\mathrm{eq}}^{\mathrm{bath}}$ is the bath equilibrium density operator at temperature $T$, i.e.~, $\hat{\rho }_{\mathrm{eq}}^{\mathrm{bath}}=e^{-\beta \hat{H}^{\mathrm{bath}}}/\mathcal{Z}$.
Here, $\beta \equiv 1/k_{\mathrm{B}}T$ is the inverse temperature divided by the Boltzmann constant $k_{\mathrm{B}}$ and $\mathcal{Z}$ is the partition function of the bath.
Note that this factorized initial condition Eq.~\eqref{eq:fact-init} is merely temporarily introduced to evaluate time evolution of the reduced system and is not a restriction of numerical calculations.
If we want to start simulations with a correlated thermal equilibrium state, we simulate time evolution of the system from temporal initial state Eq.~\eqref{eq:fact-init} to a sufficiently long time $t_{\mathrm{i}}$, and then we regard the state at $t_{\mathrm{i}}$ as a initial state of the following calculations we want.
This technique is frequently used to calculate optical response functions \cite{tanimura2006jpsj, tanimura2012jcp}.
It is also possible to obtain the correlated thermal equilibrium state as a steady--state solution of the HEOM \cite{zhang2017jcp} or as an inverse temperature integration of the imaginary-time HEOM \cite{tanimura2014jcp, tanimura2015jcp}.

The reduced system density operator at $t$,
\begin{align}
  \hat{\rho }(t)\equiv \mathrm{Tr}_{\mathrm{bath}}\{\hat{\rho }^{\mathrm{tot}}(t)\},
\end{align}
is expressed in the interaction picture as
\begin{align}
  \tilde{\rho }(t)=\mathcal{U}(t,t_{0})\tilde{\rho }(t_{0})\equiv \left\langle \mathcal{T}_{+}\exp \left(-\int _{t_{0}}^{t}\!ds\,\tilde{\mathcal{L}}^{\mathrm{int}}(s)\right)\right\rangle _{\mathrm{bath}}\tilde{\rho }(t_{0}),
  \label{eq:reduced-int}
\end{align}
where $\langle \dots \rangle _{\mathrm{bath}}\equiv \mathrm{Tr}_{\mathrm{bath}}\{\dots \rho _{\mathrm{eq}}^{\mathrm{bath}}\}$.
Because of the Gaussian property of the coordinate operator $\tilde{x}_{\xi }(s)$ in $\tilde{\mathcal{L}}^{\mathrm{int}}(s)$ via Wick's theorem, the above propagator $\mathcal{U}(t,t_{0})$ can be rewritten in the form of the second--order cumulant expansion as \cite{weiss2011book, ishizaki2009jcp}
\begin{align}
  \begin{split}
    \mathcal{U}(t,t_{0})&=\mathcal{T}_{+}\exp \Biggl[\int _{t_{0}}^{t}\!ds\,\frac{i}{\hbar }\tilde{V}(s)^{\times }\\
    &\quad \times \int _{t_{0}}^{s}\!du\,\frac{i}{\hbar }\left(\mathcal{C}(s-u)\tilde{V}(u)^{\rightarrow }-\mathcal{C}^{\ast }(s-u)\tilde{V}(u)^{\leftarrow }\right)\Biggr],
  \end{split}
\end{align}
where
\begin{align}
  \mathcal{C}(t)\equiv \langle \tilde{X}(t)\tilde{X}(0)\rangle _{\mathrm{bath}}
\end{align}
is the quantum correlation function of the collective environment coordinate $\hat{X}\equiv \sum _{\xi }g_{\xi }\hat{x}_{\xi }$, and we have introduced superoperators $\hat{A}^{\rightarrow }\hat{B}\equiv \hat{A}\hat{B}$, $\hat{A}^{\leftarrow }\hat{B}\equiv \hat{B}\hat{A}$, $\hat{A}^{\times }\hat{B}\equiv \hat{A}\hat{B}-\hat{B}\hat{A}$, and $\hat{A}^{\circ }\hat{B}\equiv \hat{A}\hat{B}+\hat{B}\hat{A}$.
Thus, the effects of the system--bath interaction is characterized by the second--order cumulant, i.e.,~$\mathcal{C}(t)$.
This correlation function is connected to the spectral density, $\mathcal{J}(\omega )\equiv \pi \sum _{\xi }g_{\xi }^{2}\delta (\omega -\omega _{\xi })/2$, as
\begin{align}
  \mathcal{C}(t)&=\frac{1}{\pi }\int _{-\infty }^{\infty }\!d\omega \,\mathcal{J}(\omega )\left(n_{\mathrm{BE}}(\omega ,T)+1\right)e^{-i\omega t}.
  \label{eq:C}
\end{align}
Here, $n_{\mathrm{BE}}(\omega ,T)=(e^{\beta \hbar \omega }-1)^{-1}$ is the Bose--Einstein distribution function.
The real and imaginary parts of the quantum correlation function, i.e., the symmetrized correlation function $\mathcal{S}(t)\equiv (\mathcal{C}(t)+\mathcal{C}^{\ast }(t))/2$ and anti--symmetrized correlation function $\mathcal{A}(t)\equiv (\mathcal{C}(t)-\mathcal{C}^{\ast }(t))/2i$, represent fluctuation and dissipation of the bath, respectively.
Note that
\begin{subequations}
  \begin{align}
    \mathcal{S}(t)&=\frac{2}{\pi }\int _{0}^{\infty }\!d\omega \,\mathcal{J}(\omega )\left(n_{\mathrm{BE}}(\omega ,T)+\frac{1}{2}\right)\cos \omega t
    \label{eq:S}\\
    \intertext{and}
    \mathcal{A}(t)&=-\frac{1}{\pi }\int _{0}^{\infty }\!d\omega \,\mathcal{J}(\omega )\sin \omega t
    \label{eq:A}.
  \end{align}
\end{subequations}

By using $\mathcal{S}(t)$ and $\mathcal{A}(t)$, the propagator can be rewritten as
\begin{align}
  \mathcal{U}(t,t_{0})&=\mathcal{T}_{+}\mathcal{F}(t,t_{0})\\
  \intertext{and}
  \mathcal{F}(t,t_{0})&\equiv \exp \Biggl[\int _{t_{0}}^{t}\!ds\,\tilde{\Phi }(s)\notag\\
    &\quad \quad \times \int _{t_{0}}^{s}\!du\,\left(\mathcal{S}(s-u)\tilde{\Phi }(u)-\mathcal{A}(s-u)\tilde{\Psi }(u)\right)\Biggr]
  \label{eq:cumulant}.
\end{align}
Here, $\hat{\Phi }\equiv i\tilde{V}^{\times }/\hbar $ and $\hat{\Psi }\equiv \tilde{V}^{\circ }/\hbar $.

\subsection{Hierarchical equations of motion}
We prepare a set of $K$ basis functions of time $t$, ${}^{t}\bm{\phi }(t)=(\dots ,\phi _{k}(t),\dots )$, which satisfy a set of time evolution equations
\begin{align}
  \partial _{t}\bm{\phi }(t)=-\bm{\gamma }\bm{\phi }(t).
  \label{eq:basis-evolution}
\end{align}
Here, the superscripts ${}^{t}$ on matrix/vector represent the transposes of the matrix/vector, and $(\bm{\gamma })_{jk}=\gamma _{jk}$ is a $K\times K$ complex matrix which can be non--diagonalizable.
Note that, to obtain stable time evolution of the basis $\bm{\phi }(t)$, the real parts of non--degenerate and degenerate eigenvalues of the coefficient matrix $\bm{\gamma }$ should be non-negative and positive, respectively.
To construct the ``hierarchy'' later, we need to express the symmetrized and anti--symmetrized correlation functions as linear combinations of $\bm{\phi }(t)$ and the Dirac delta function $\delta (t)$\footnote{ In this paper, we treat $\delta (t)$ as an even function, and therefore $\int _{0}^{\infty }\!dx\,f(x)\delta (x)=f(0)/2$.
}, i.e., $\mathcal{S}(t)=\sum _{k}S_{k}\cdot \phi _{k}(t)+S_{\delta }\cdot 2\delta (t)={}^{t}\bm{S}\bm{\phi }(t)+S_{\delta }\cdot 2\delta (t)$ and $\mathcal{A}(t)=\sum _{k}A_{k}\cdot \phi _{k}(t)={}^{t}\bm{A}\bm{\phi }(t)$.
In this paper, we assume that the symmetrized and anti--symmetrized correlation functions can be rewritten as
\begin{subequations}
  \begin{align}
    &\begin{aligned}
      \mathcal{S}(t)&=\sum _{jk}\sigma _{j}s_{jk}\phi _{k}(t)+S_{\delta }\cdot 2\delta (t)\\
      &={}^{t}\bm{\sigma }\bm{s}\bm{\phi }(t)+S_{\delta }\cdot 2\delta (t)
      \label{eq:S-expansion}
    \end{aligned}
    \intertext{and}
    &\begin{aligned}
      \mathcal{A}(t)&=\sum _{jk}\sigma _{j}a_{jk}\phi _{k}(t)\\
      &={}^{t}\bm{\sigma }\bm{a}\bm{\phi }(t),
      \label{eq:A-expansion}
    \end{aligned}
  \end{align}
\end{subequations}
i.e.,
\begin{align}
  {}^{t}\bm{S}={}^{t}\bm{\sigma }\bm{s}~\text{and}~{}^{t}\bm{A}={}^{t}\bm{\sigma }\bm{a}.
  \label{eq:S-s-relation}
\end{align}
Here, ${}^{t}\bm{\sigma }=(\dots ,\sigma _{k},\dots )$ is a constant vector which is common in $\mathcal{S}(t)$ and $\mathcal{A}(t)$, and $\bm{s}$ and $\bm{a}$ are $K\times K$ complex matrices, which commute with $\bm{\gamma }$.
Note that the parametrization of $\bm{\sigma }$, $\bm{s}$, and $\bm{a}$ is not unique.
In Appendix \ref{sec:commuting-s-a}, we give examples of possible constructions of $\bm{\sigma }$, $\bm{s}$, and $\bm{a}$ form $\bm{S}$ and $\bm{A}$.
By substituting Eqs.~\eqref{eq:S-expansion} and \eqref{eq:A-expansion} into cumulant expansion Eq.~\eqref{eq:cumulant},  we get
\begin{align}
  \mathcal{F}(t,t_{0})&=\exp \Biggl[\int _{t_{0}}^{t}\!ds\,\biggl(-\tilde{\Xi }(s)+\sum _{j}\tilde{\Phi }_{j}(s)\int _{t_{0}}^{s}\!du\,\tilde{\Theta }_{j}(s,u)\biggr)\Biggr],
\end{align}
where
\begin{subequations}
  \begin{align}
    \tilde{\Xi }(s)&\equiv -S_{\delta }\tilde{\Phi }(s)^{2}
  \end{align}
  is describing the effect of Markovian part of the bath correlation function, and
  \begin{align}
    \tilde{\Phi }_{j}(s)&\equiv \sigma _{j}\tilde{\Phi }(s)\\
    \intertext{and}
    \tilde{\Theta }_{j}(t,s)&\equiv \sum _{k}\left(s_{jk}\phi _{k}(t-s)\tilde{\Phi }(s)-a_{jk}\phi _{k}(t-s)\tilde{\Psi }(s)\right)
  \end{align}
  are describing non--Markovian effects by delay functions $\{\phi _{k}(t)\}$.
\end{subequations}
Because $\bm{s}$ and $\bm{a}$ commute with $\bm{\gamma }$, $\Theta _{k}(t,s)$ satisfies a time evolution equation similar to Eq.~\eqref{eq:basis-evolution} as follows:
\begin{widetext}
  \begin{align}
    \begin{split}
      \partial _{t}\tilde{\Theta }_{j}(t,s)&=-\sum _{k}\sum _{l}\left(s_{jk}\gamma _{kl}\phi _{l}(t-s)\tilde{\Phi }(s)-a_{jk}\gamma _{kl}\phi _{l}(t-s)\tilde{\Psi }(s)\right)\\
      &=-\sum _{k}\gamma _{jk}\sum _{l}\left(s_{kl}\phi _{l}(t-s)\tilde{\Phi }(s)-a_{kl}\phi _{l}(t-s)\tilde{\Psi }(s)\right)\\
      &=-\sum _{k}\gamma _{jk}\tilde{\Theta }_{k}(t,s).
    \end{split}
  \end{align}
\end{widetext}
This result is the key point of our new treatment.

To evaluate the time evolution of the reduced density operator $\hat{\rho }(t)$, we introduce the auxiliary density operators (ADOs) defined by
\begin{align}
  \begin{split}
    \tilde{\rho }_{\bm{n}}(t)&\equiv \mathcal{T}_{+}\prod _{k}\left(-\int _{t_{0}}^{t}\!ds\,\tilde{\Theta }_{k}(t,s)\right)^{n_{k}}\mathcal{F}(t,t_{0})\tilde{\rho }(t_{0}),
    \label{eq:ado}
  \end{split}
\end{align}
where $\bm{n}=(\dots ,n_{k},\dots )$ is a $K$--dimensional multi--index whose components are non--negative integers.
Cleary, $\tilde{\rho }_{\bm{0}}(t)$ equals to the reduced density operator $\tilde{\rho }(t)$, i.e., Eq.~\eqref{eq:reduced-int}.
By calculating first--order time derivative of Eq.~\eqref{eq:ado} with respect to time $t$, we obtain a set of time evolution equations in the interaction picture as
\begin{align}
  \begin{split}
    \partial _{t}\tilde{\rho }_{\bm{n}}(t)&=-\tilde{\Xi }(t)\tilde{\rho }_{\bm{n}}(t)-\sum _{j,k}n_{j}\gamma _{jk}\tilde{\rho }_{\bm{n}-\bm{1}_{j}+\bm{1}_{k}}(t)\\
    &\quad -\sum _{k}\tilde{\Phi }_{k}(t)\tilde{\rho }_{\bm{n}+\bm{1}_{k}}(t)-\sum _{k}n_{k}\tilde{\Theta }_{k}(t,t)\tilde{\rho }_{\bm{n}-\bm{1}_{k}}(t).
  \end{split}
  \label{eq:heom-int}
\end{align}
Here, $\bm{1}_{k}=(0,\dots ,1,0,\dots )$ is the $k$th unit vector.
These can be rewritten in the Schr\"odinger picture as
\begin{align}
  \begin{split}
    \partial _{t}\hat{\rho }_{\bm{n}}(t)&=-(\mathcal{L}+\hat{\Xi })\hat{\rho }_{\bm{n}}(t)-\sum _{j,k}n_{j}\gamma _{jk}\hat{\rho }_{\bm{n}-\bm{1}_{j}+\bm{1}_{k}}(t)\\
    &\quad -\sum _{k}\hat{\Phi }_{k}\hat{\rho }_{\bm{n}+\bm{1}_{k}}(t)-\sum _{k}n_{k}\hat{\Theta }_{k}\hat{\rho }_{\bm{n}-\bm{1}_{k}}(t).
    \label{eq:heom}
  \end{split}
\end{align}
Here,
\begin{subequations}
  \begin{align}
    \mathcal{L}&\equiv -\frac{i}{\hbar }\hat{H}^{\times },\quad \hat{\Xi }\equiv -S_{\delta }\hat{\Phi }^{2},\quad \hat{\Phi }_{k}\equiv \sigma _{k}\hat{\Phi },\\
    \intertext{and}
    \hat{\Theta }_{k}&\equiv \sum _{l}s_{kl}\phi _{l}(0)\hat{\Phi }-\sum _{l}a_{kl}\phi _{l}(0)\hat{\Psi }.
  \end{align}
\end{subequations}
This is our generalization of the HEOM.

\begin{figure}
  \centering
  \includegraphics[scale=\SingleColFigScale]{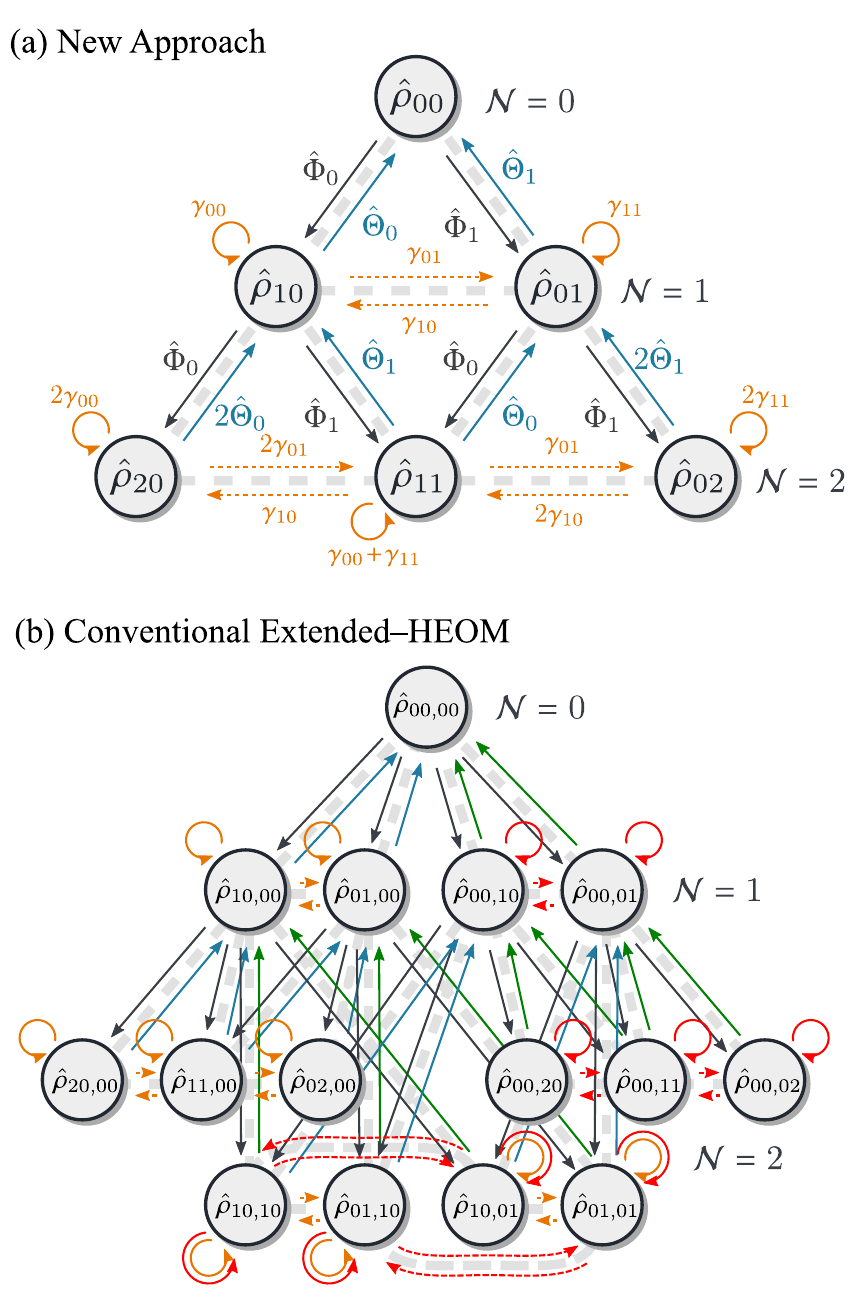}
  \caption{ (a) The hierarchical structure of Eq.~\eqref{eq:heom} in the case of $K=2$ and $\mathcal{N}_{\mathrm{max}}=2$.
    Here, the first term in Eq.~\eqref{eq:heom} is omitted.
    Starting and end points of arrows indicate the left--hand side and right--hand side terms of Eq.~\eqref{eq:heom}.
    Orange, black, and blue arrows represent the second, third, and fourth terms.
    The dashed orange arrows refer to connections caused by off--diagonal elements of $\bm{\gamma }$, which do not exist in the original formulation.
    (b) One of the conventional generalizations, the extended HEOM Eq.~\eqref{eq:eheom}, in the case of $K=2$ and $\mathcal{N}_{\mathrm{max}}=2$.
    Orange, red, black, blue, and green arrows represent the second, third, fourth, fifth, and sixth terms of Eq.~\eqref{eq:eheom}, respectively.
  }
  \label{fig:hierarchy}
\end{figure}
In Fig.~\ref{fig:hierarchy}(a), a schematic structure of connections of ADOs in Eq.~\eqref{eq:heom} is depicted.
The set of equations forms a hierarchical structure, of which top element is the reduced density operator $\hat{\rho }_{\bm{0}}(t)=\hat{\rho }(t)$.
The first term of Eq.~\eqref{eq:heom} represents Markovian dynamics caused by the free propagation of the system and the Markovian part of the bath.
The second term means time evolution of the basis set $\bm{\phi }(t)$, which expresses non--Markovian free propagation of the bath.
The third and fourth terms represent system--environment interactions, which include the effects of fluctuation and dissipation of the bath via $\mathcal{S}(t)$ and $\mathcal{A}(t)$.

Because Eq.~\eqref{eq:heom} consists of an infinite number of differential equations, we need to truncate $\bm{n}$ to carry out numerical calculations.
The method of truncation affects the efficiency of calculations, and many advanced methods are proposed within the original HEOM framework \cite{tanimura1994jpsj, ishizaki2005jpsj, shi2009jcp3, hartle2013prb, hartle2015prb}.
In this article, in order to make the validation of theories simple, we truncate them in accordance with the condition that $\bm{n}$ satisfies the relation $\mathcal{N}\equiv \sum _{k}n_{k}>\mathcal{N}_{\mathrm{max}}$, and we regard the ADOs which satisfy $\mathcal{N}>\mathcal{N}_{\mathrm{max}}$ as zero.
Here, $\mathcal{N}$ is referred to as a tier of an ADO.
In this manner, we need to judge convergence of the numerical results by changing $\mathcal{N}_{\mathrm{max}}$.

In numerical implementations, it is more convenient to define ADOs by $\hat{\rho }'_{\bm{n}}=\hat{\rho }_{\bm{n}}/\prod _{k}\sqrt {\mathstrut n_{k}!}$ to suppress divergence of ADOs in deep tiers $\mathcal{N}\gg 1$ in the case of strong system--bath coupling \cite{shi2009jcp3}.
Then the equations of motion are rewritten as
\begin{align}
  \begin{split}
    \partial _{t}\hat{\rho }'_{\bm{n}}(t)&=-(\mathcal{L}+\hat{\Xi })\hat{\rho }_{\bm{n}}'(t)-\sum _{k}n_{j}\gamma _{kk}\hat{\rho }_{\bm{n}}'(t)\\
    &\quad -\sum _{j\neq k}\sqrt {n_{j}}\sqrt {\mathstrut n_{k}+1}\gamma _{jk}\hat{\rho }_{\bm{n}-\bm{1}_{j}+\bm{1}_{k}}'(t)\\
    &\quad -\sum _{k}\sqrt {\mathstrut n_{k}+1}\hat{\Phi }_{k}\hat{\rho }_{\bm{n}+\bm{1}_{k}}'(t)-\sum _{k}\sqrt {\mathstrut n_{k}}\hat{\Theta }_{k}\hat{\rho }_{\bm{n}-\bm{1}_{k}}'(t).
    \label{eq:heom-prime}
  \end{split}
\end{align}

\subsection{Relation to conventional theories}
In this section, we show the relation among our new generalization, the original theory, and extended HEOM.
The relation among our approach and some other generalizations of HEOM is given in Appendix~\ref{eq:rel-yheom}.

\subsubsection{Reduction to the original HEOM}
\label{sec:rel-oheom}
When the coefficient matrix $\bm{\gamma }$ has only diagonal elements (i.e., $\gamma _{kk}\equiv \gamma _{k}$), Eq.~\eqref{eq:basis-evolution} is solved as
\begin{align}
  \phi _{k}(t)=e^{-\gamma _{k}t} &&(t\geq 0).
\end{align}
Here, we have set $\bm{\phi }(0)$ as $\phi _{k}(0)=1$.
In this case, $\bm{s}$ and $\bm{a}$, which commute with $\bm{\gamma }$, should also have only diagonal elements (i.e., $s_{kk}\equiv s_{k}$ and $a_{kk}\equiv a_{k}$), and $\mathcal{S}(t)$ and $\mathcal{A}(t)$ can be rewritten as $\mathcal{S}(t)=\sum _{k}s_{k}\cdot e^{-\gamma _{k}\left|t\right|}+S_{\delta }\cdot 2\delta (t)$ and $\mathcal{A}(t)=\sum _{k}a_{k}\cdot e^{-\gamma _{k}\left|t\right|}$ (We have fixed the parameter $\bm{\sigma }$ as $\sigma _{k}=1$).
As a result, Eq.~\eqref{eq:heom} reduces to
\begin{align}~
  \begin{split}
    \partial _{t}\hat{\rho }_{\bm{n}}(t)&=-\left(\mathcal{L}+\hat{\Xi }+\sum _{k}n_{k}\gamma _{k}\right)\hat{\rho }_{\bm{n}}(t)\\
    &\quad -\sum _{k}\hat{\Phi }\hat{\rho }_{\bm{n}+\bm{1}_{k}}(t)-\sum _{k}n_{k}\hat{\Theta }_{k}\hat{\rho }_{\bm{n}-\bm{1}_{k}}(t).
    \label{eq:oheom}
  \end{split}
\end{align}
Here,
\begin{align}
  \hat{\Theta }_{k}&\equiv s_{k}\hat{\Phi }-a_{k}\hat{\Psi }.
\end{align}
This is the original HEOM.
Thus, the original HEOM can be regarded as a special case of our generalization.

\subsubsection{Relation to the extended HEOM}
\label{sec:rel-eheom}
In the extended HEOM approach, we need to duplicate the basis $\bm{\phi }(t)$ as ${}^{t}\bm{\phi }'(t)=({}^{t}\bm{\phi }(t), {}^{t}\bm{\phi }(t))$.
This satisfies time evolution
\begin{align}
  \partial _{t}
  \bm{\phi }'(t)
  &=
  -\bm{\gamma }'\bm{\phi }'(t)
  \equiv 
  -\begin{pmatrix}
  \bm{\gamma }&\bm{0}\\
  \bm{0}&\bm{\gamma }\\
  \end{pmatrix}
  \begin{pmatrix}
    \bm{\phi }(t)\\
    \bm{\phi }(t)\\
  \end{pmatrix}.
  \label{eq:gamma_prime}
\end{align}
Expansion of correlation functions $\mathcal{S}(t)$ and $\mathcal{A}(t)$ for the extended HEOM can be written in the form of Eqs.~\eqref{eq:S-expansion} and \eqref{eq:A-expansion} as
\begin{subequations}
  \begin{align}
    &\begin{aligned}
       \mathcal{S}(t)&=
       {}^{t}\bm{\sigma }'\bm{s}'\bm{\phi }'(t)\equiv 
       \begin{pmatrix}
         {}^{t}\bm{\sigma }_{\mathcal{S}}&{}^{t}\bm{\sigma }_{\mathcal{A}}\\
       \end{pmatrix}
       \begin{pmatrix}
         \bm{1}&\bm{0}\\
         \bm{0}&\bm{0}\\
       \end{pmatrix}
       \begin{pmatrix}
         \bm{\phi }(t)\\
         \bm{\phi }(t)\\
       \end{pmatrix}\\
       &=\sum _{k}\sigma _{\mathcal{S},k}\phi _{k}(t)
     \end{aligned}
    \intertext{and}
    &\begin{aligned}
       \mathcal{A}(t)&=
       {}^{t}\bm{\sigma }'\bm{a}'\bm{\phi }'(t)\equiv 
      \begin{pmatrix}
        {}^{t}\bm{\sigma }_{\mathcal{S}}&{}^{t}\bm{\sigma }_{\mathcal{A}}\\
      \end{pmatrix}
      \begin{pmatrix}
        \bm{0}&\bm{0}\\
        \bm{0}&\bm{1}\\
      \end{pmatrix}
      \begin{pmatrix}
        \bm{\phi }(t)\\
        \bm{\phi }(t)\\
      \end{pmatrix}\\
      &=\sum _{k}\sigma _{\mathcal{A},k}\phi _{k}(t)
    \end{aligned}
  \end{align}
\end{subequations}
Here, $\bm{0}$ and $\bm{1}$ are the $K\times K$ zero and identity matrices, respectively, and we have assumed that $S_{\delta }=0$.
Obviously, $\sigma _{\mathcal{S},k}=S_{k}$ and $\sigma _{\mathcal{A},k}=A_{k}$, and $\bm{a}'$ and $\bm{s}'$ commute with $\bm{\gamma }'$.
Then the number of the indices of the ADOs are also doubled as $\bm{n}=(\bm{l},\bm{m})$, and we get the corresponding HEOM from Eq.~\eqref{eq:heom} as
\begin{widetext}
  \begin{align}
    \begin{split}
      \partial _{t}\hat{\rho }_{\bm{l},\bm{m}}(t)&=-\mathcal{L}\hat{\rho }_{\bm{l},\bm{m}}(t)\\
      &\quad -\sum _{j,k}\gamma _{jk}l_{j}\hat{\rho }_{\bm{l}-\bm{1}_{j}+\bm{1}_{k},\bm{m}}(t)-\sum _{j,k}\gamma _{jk}m_{j}\hat{\rho }_{\bm{l},\bm{m}-\bm{1}_{j}+\bm{1}_{k}}(t)\\
      &\quad -\sum _{k}\hat{\Phi }\left(\sigma _{\mathcal{S},k}\hat{\rho }_{\bm{l}+\bm{1}_{k},\bm{m}}(t)+\sigma _{\mathcal{A},k}\hat{\rho }_{\bm{l},\bm{m}+\bm{1}_{k}}(t)\right)\\
      &\quad -\sum _{k}\phi _{k}(0)l_{k}\hat{\Phi }\hat{\rho }_{\bm{l}-\bm{1}_{k},\bm{m}}(t)+\sum _{k}\phi _{k}(0)m_{k}\hat{\Psi }\hat{\rho }_{\bm{l},\bm{m}-\bm{1}_{k}}(t).
      \label{eq:eheom}
    \end{split}
  \end{align}
\end{widetext}
This is equivalent to the extended HEOM.
As shown here, the extended HEOM can be regarded as a subset of our new treatment.

In Fig.~\ref{fig:hierarchy}(b), a schematic structure of connections of ADOs in Eq.~\eqref{eq:eheom} is depicted within the same condition of Fig.~\ref{fig:hierarchy}(a).
In comparison with the extended HEOM, our new approach displays a simple hierarchical structure without doubling the number of indices of the hierarchy, and the simple structure has huge theoretical and numerical advantages, especially when $\mathcal{K}$ and $\mathcal{N}_{\mathrm{max}}$ are large.

\section{NUMERICAL RESULTS}
\label{sec:results}
Our generalization Eq.~\eqref{eq:heom} works for arbitrary spectral densities at arbitrary temperature as far as accurate expansions bath correlation functions Eqs.~\eqref{eq:S-expansion} and \eqref{eq:A-expansion} are given and the required ADOs converge in a finite number.
In this section, we demonstrate our new treatment for non--exponential correlation functions by using three examples.
Hereafter, we employ the dimensionless units $\hbar =1$ and $k_{\mathrm{B}}=1$ for simplicity.
Numerical calculations were carried out to integrate time differential equations by using the fourth--order low--storage Runge--Kutta (LSRK4) method \cite{williamson1980jcp}.
The time step for the LSRK4 method was set to $\delta t=10^{-2}$.

\subsection{Super--Ohmic spectral density with Bessel function basis}
\label{sec:example1}
\begin{figure}
  \centering
  \includegraphics[scale=\SingleColFigScale]{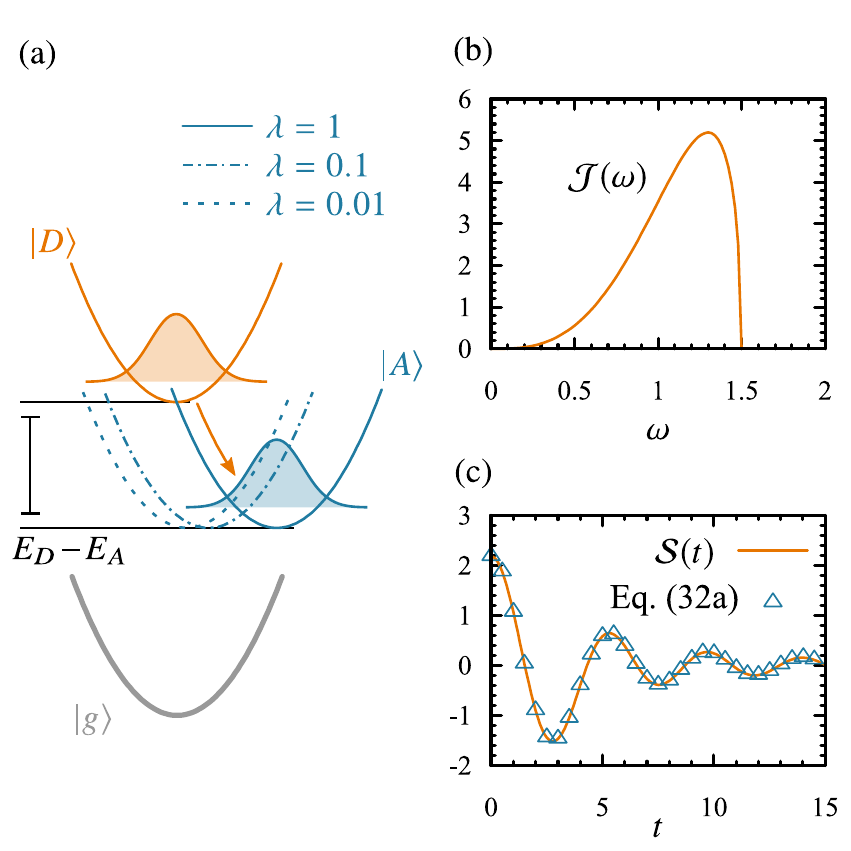}
  \caption{ (a) Donor--acceptor model system of an exciton/electron transfer problem in this article.
    Gray, orange, and blue curves represent free energy surfaces of $|g\rangle $, $|D\rangle $, and $|A\rangle $, respectively.
    Here, $E_{\mathrm{D}}-\mathrm{E}_{A}=1$, and in the cases of $\lambda =0.01$, $0.1$, $1$ are depicted.
    (b) Super--Ohmic spectral density model with a semicircle cutoff function, Eq.~\eqref{eq:semicircle}.
    Here, $\lambda =1$ and $\gamma _{\mathrm{c}}=1.5$.
    (c) The symmetrized correlation function of the spectral density Eq.~\eqref{eq:semicircle}.
    Blue triangles represent results of the expansion Eq.~\eqref{eq:S-bessel} with the truncated basis set $\phi (t)$.
    Here, the parameters of the spectral density are the same as (b), and the other parameters are $T=1$ and $K=16$.
  }
  \label{fig:semicircle}
\end{figure}
In this section, we consider exciton/electron transfer problems in donor--acceptor system with a super--Ohmic spectral density.
We consider a three--level system which has ground state $|g\rangle $, reactant state of the transfer $|D\rangle $ (donor state), and product state of the transfer $|A\rangle $ (acceptor state).
In the case of an electron transfer problem, the states are, e.g., $\mathrm{D}\mathrm{A}$, $\mathrm{D}^{\ast }\mathrm{A}$, and $\mathrm{D}^{+}\mathrm{A}^{-}$, respectively.
Here, $\mathrm{D}$ is a donor molecule and $\mathrm{A}$ is an acceptor molecule.
Typically, degrees of freedom of reorganization after the transitions $|g\rangle \rightarrow |D\rangle $ and $|D\rangle \rightarrow |A\rangle $ are different, and therefore we need to consider two sets of phonon degrees of freedom to describe them.
In this article, to make our demonstration simple, we assume that the transition $|g\rangle \rightarrow |D\rangle $ does not cause reorganization of the molecule (i.e., the stable points on $|g\rangle $ and $|D\rangle $ are the same) as depicted in Fig.~\ref{fig:semicircle}(a).
Then, we need only a single set of phonon degrees of freedom, and the total Hamiltonian can be expressed as \cite{yang2002cp}
\begin{align}
  \begin{split}
    \hat{H}^{\mathrm{tot}}&=E_{g}|g\rangle \langle g|+E_{D}|D\rangle \langle D|+E_{A}|A\rangle \langle A|\\
    &\quad +J(|D\rangle \langle A|+|A\rangle \langle D|)\\
    &\quad +\left(|g\rangle \langle g|+|D\rangle \langle D|\right)\sum _{\xi }\frac{\hbar \omega _{\xi }}{2}\left(\hat{p}_{\xi }^{2}+\hat{x}_{\xi }^{2}\right)\\
    &\quad +|A\rangle \langle A|\sum _{\xi }\frac{\hbar \omega _{\xi }}{2}\left[\hat{p}_{\xi }^{2}+\left(\hat{x}_{\xi }-\frac{g_{\xi }}{\hbar \omega _{\xi }}\right)^{2}\right].
    \label{eq:hamiltonian-et}
  \end{split}
\end{align}
Here, $E_{g}$, $E_{D}$, and $E_{A}$ are stable equilibrium energies of each state, and $J$ is the electronic coupling between donor and acceptor states.
Spontaneous transitions between $|g\rangle $ and $\{|D\rangle , |A\rangle \}$ are not considered.
The displacement $g_{\xi }/\hbar \omega _{\xi }$ represents the difference of the stable points between $|D\rangle $ and $|A\rangle $ in the $\xi $th phonon degree of freedom, and total reorganization energy after $|D\rangle \rightarrow |A\rangle $ transition can be written as $\lambda =g_{\xi }^{2}/2\hbar \omega _{\xi }$.
As far as we consider transfer problem after excitation $|g\rangle \rightarrow |D\rangle $, we can ristrict the space of the problem in $\{|D\rangle , |A\rangle \}$, and then Eq.~\eqref{eq:hamiltonian-et} can be rewritten in the form of Eq.~\eqref{eq:total-hamiltonian} with $\hat{H}=E_{\mathrm{D}}|D\rangle \langle D|+(E_{A}+\lambda )|A\rangle \langle A|+J(|D\rangle \langle A|+|A\rangle \langle D|)$ and $\hat{V}=|A\rangle \langle A|$.
Here, we adopt a super--Ohmic spectral density model with a semicircle cutoff function,
\begin{align}
  \mathcal{J}(\omega )=
  \begin{cases}
    \displaystyle \frac{16\lambda }{\gamma _{\mathrm{c}}^{3}}\omega ^{3}\sqrt {1-\omega ^{2}/\gamma _{\mathrm{c}}^{2}}&(\left|\omega \right|\leq \gamma _{\mathrm{c}})\\
    0,&(\left|\omega \right|>\gamma _{\mathrm{c}})
  \end{cases}
  \label{eq:semicircle}
\end{align}
which is proposed to describe electron transfer in bacterial photosynthesis \cite{ando1998jpcb}.
Here, $\gamma _{\mathrm{c}}$ is the cutoff frequency of the phonon.
In this spectral density model, the density has a rigid cutoff at $\omega =\gamma _{\mathrm{c}}$ as depicted in Fig.~\ref{fig:semicircle}(b), and this describes the upper--limit of the frequencies in a molecule--environment system, which should have a finite value.
In the case of this spectral density model, the correlation functions Eqs.~\eqref{eq:S} and \eqref{eq:A} are analytically evaluated in the forms of linear combinations the Bessel functions of the first kind $J_{k}(x)$ as
\begin{subequations}
  \begin{align}
    \mathcal{S}(t)&=\sum _{k=0}S_{k}\cdot J_{k}(\gamma _{\mathrm{c}}t)
    \label{eq:S-bessel}
    \intertext{and}
    \mathcal{A}(t)&=\sum _{k=0}A_{k}\cdot J_{k}(\gamma _{\mathrm{c}}t)
    \label{eq:A-bessel}.
  \end{align}
\end{subequations}
For details of the coefficients, see Appendix \ref{sec:decomposition-superohmic}.
Thus, we can employ the Bessel functions as the basis functions for HEOM, $\phi _{k}(t)=J_{k}(\gamma _{\mathrm{c}}t)$, which satisfy time evolution equations
\begin{subequations}
  \begin{align}
    \partial _{t}\phi _{0}(t)&=+\gamma _{\mathrm{c}}\phi _{1}(t)\\
    \intertext{and}
    \partial _{t}\phi _{k}(t)&=-\gamma _{\mathrm{c}}\phi _{k-1}(t)/2+\gamma _{\mathrm{c}}\phi _{k+1}(t)/2&(k\geq 1).
    \label{eq:bessel-basis}
  \end{align}
\end{subequations}
These are an infinite number of simultaneous equations, so we need to truncate them in a finite number $K$.
Although this truncation introduces an approximation in HEOM calculations, we can test the accuracy of the results by changing $K$  and we can obtain solution as accurate as we need.
In the calculations in this section, we simply ignore $\phi _{K}(t)$.
Note that, this truncation also makes long time behavior of the basis functions unstable.
Therefore we need to increase $K$ when we want to increase the simulation times of our calculations.

We set the parameters as $E_{\mathrm{D}}-E_{\mathrm{A}}=1$, $J=0.5$, $\gamma _{\mathrm{c}}=1.5$, and $T=1$.
The truncation number of Eq.~\eqref{eq:bessel-basis} was chosen to be $K=15$, which sufficiently reproduces analytical correlation functions in our simulation time $0\leq t\leq 15$ as depicted in Fig.~\ref{fig:semicircle}(c).
Because $\bm{\gamma }$ from Eq.~\eqref{eq:bessel-basis} is diagonalizable, we constructed $\bm{\sigma }$, $\bm{s}$, and $\bm{a}$ by using the method shown in Appendix \ref{sec:commuting-s-a-1}

\begin{figure}
  \centering
  \includegraphics[scale=\SingleColFigScale]{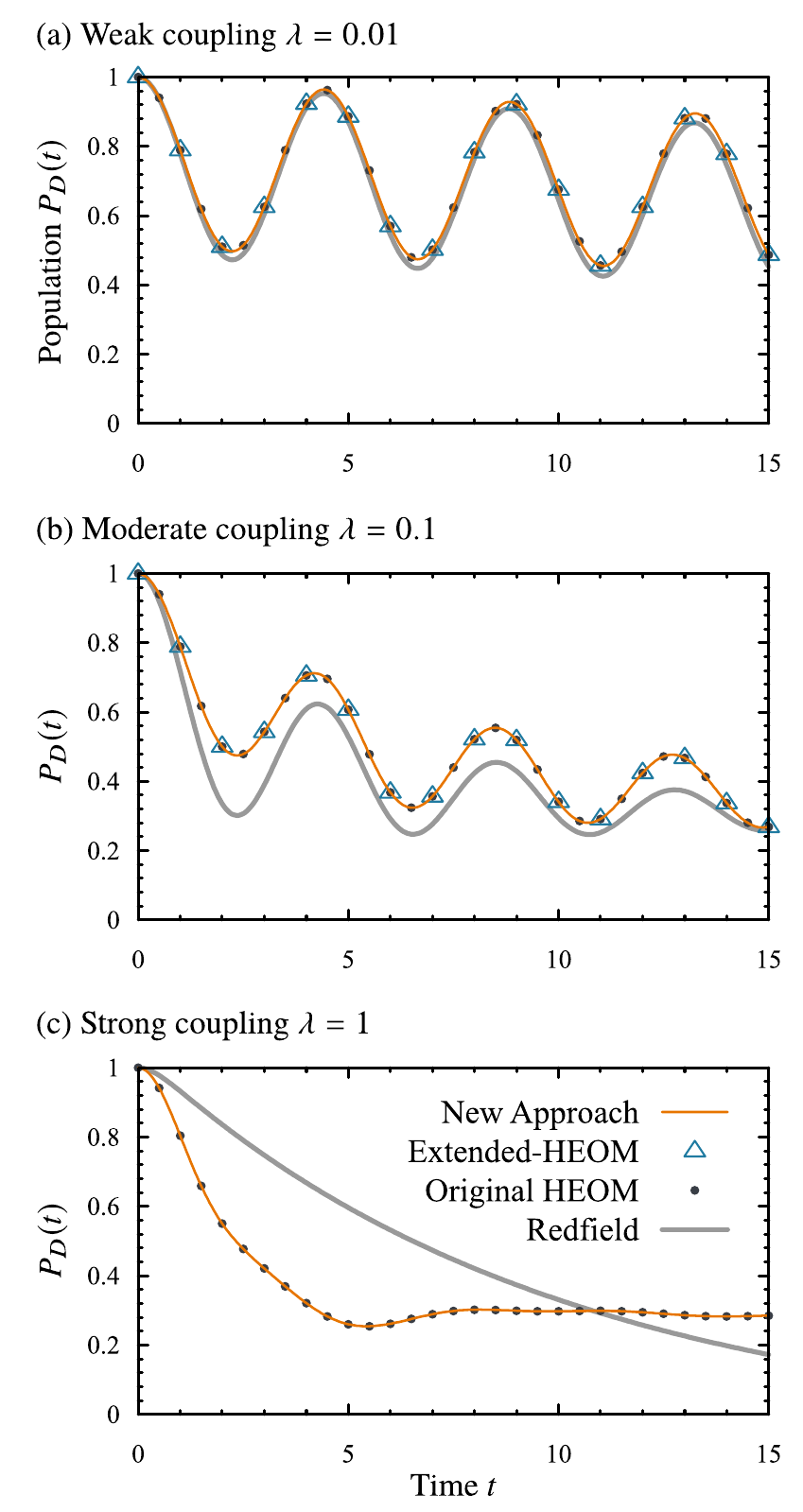}
  \caption{ Donor population dynamics after excitation at $t=0$ in the case of (a) weak coupling (reorganization energy) $\lambda =0.01$, (b) moderate coupling $\lambda =0.1$, and (c) strong coupling $\lambda =1$.
    Orange curves, blue triangles, and black dots represent results of new treatment (Eq.~\eqref{eq:heom}), extended HEOM (Eq.~\eqref{eq:eheom}), and original HEOM (Eq.~\eqref{eq:oheom}) with diagonalized $\bm{\gamma }$.
    Gray curves represent results of the Redfield equation without the secular approximation. }
  \label{fig:superohmic}
\end{figure}
In Fig.~\ref{fig:superohmic}, donor populations as a function of time $t$ are depicted for (a) weak, moderate, and strong coupling cases ($\lambda =0.01$, $0.1$, and $1$, respectively).
We performed simulations by using our new treatment (Eq.~\eqref{eq:heom}), the extended HEOM, and Redfield equations \cite{redfield1965inbook, yang2002cp, ishizaki2009jcp2} for comparison.
The truncation tiers of HEOM were chosen as $\mathcal{N}_{\mathrm{max}}=3$ for weak and moderate couplings and $\mathcal{N}_{\mathrm{max}}=8$ for strong coupling.
Under this truncation, the numbers of ADOs for Eq.~\eqref{eq:heom} are $968$ and $735,470$, and those of the extended HEOM \eqref{eq:eheom} are $6,544$ and $76,904,684$.
Because the extended HEOM calculation with $\mathcal{N}_{\mathrm{max}}=8$ was numerically high--cost, we did not perform it.
As shown, our new treatment completely reproduced the results of the extended HEOM, while computational costs are much smaller.
In Fig.~\ref{fig:superohmic}(a), the results of HEOM theories and Redfield theory are with almost coincident, because the Born and Markov approximations in the Redfield theory is valid when the coupling sufficiently small.
On the other hand, in Figs.~\ref{fig:semicircle}(b) and \ref{fig:superohmic}(c), the results of Redfield theory are different from the other calculations.
This indicates that the results of our new treatment are equivalent to that of the extended HEOM beyond a perturbative Markovian regime.

As shown here, our new treatment can describe non--exponential behavior of correlation functions.
However, the coefficient matrix $\bm{\gamma }$ in this section is diagonalizable as mentioned above, and it is possible to construct an equivalent original HEOM by diagonalization (black dots in Fig.~\ref{fig:superohmic}), which is more efficient than the calculations with non--diagonalized $\bm{\gamma }$.
Generally, as far as $\bm{\gamma }$ is diagonalizable, the original framework of the HEOM with diagonalized $\bm{\gamma }$ works well, and there is no need to use the other generalizations.
The generalizations for non--exponential correlation functions make a difference when $\bm{\gamma }$ is non--diagonalizable.
In the following two sections, we show examples with non--diagonalizable $\bm{\gamma }$.

\subsection{Critically damped Brownian oscillator}
\label{sec:example2}
In this section, we show exciton/electron transfer models coupled to damped vibrational degrees of freedom, expressed by using a Brownian spectral density \cite{tanimura1994jpsj, tanaka2009jpsj}.
The model Hamiltonian and parameters are the same as those in Sec.~\ref{sec:example1}, while we employ a Brownian spectral density
\begin{align}
  \mathcal{J}(\omega )&=2\lambda \frac{\zeta \omega _{0}^{2}\omega }{(\omega ^{2}-\omega _{0}^{2})^{2}+\zeta ^{2}\omega ^{2}}
  \label{eq:brownian}
\end{align}
instead of Eq.~\eqref{eq:semicircle}.
This spectral density represents a Brownian motion of a harmonic oscillator with frequency $\omega _{0}$ and friction constant $\zeta $.
The conditions $\zeta <2\omega _{0}$, $\zeta =2\omega _{0}$, and $\zeta >2\omega _{0}$ correspond to the underdamped, critically damped, and overdamped cases of the oscillator, respectively.
When the system $\hat{H}$ is coupled to a primary harmonic mode and the mode is coupled to another Ohmic bath, the effective spectral density the mode engenders is analytically reduced to Eq.~\eqref{eq:brownian} \cite{garg1985jcp}, and this can be regarded as a simplest model of vibronic phenomena in a dissipative environment.

The anti--symmetrized correlation function Eq.~\eqref{eq:A} is evaluated as
\begin{align}
  \mathcal{A}(t)&=\frac{\lambda \omega _{0}^{2}i}{2\omega _{1}}\left(e^{-\gamma _{+}\left|t\right|}-e^{-\gamma _{-}\left|t\right|}\right),
\end{align}
where $\omega _{1}\equiv \sqrt {\mathstrut \omega _{0}^{2}-\zeta ^{2}/4}$ and $\gamma _{\pm }\equiv \zeta /2\mp i\omega _{1}$, and typicall we employ $\phi _{+}(t)\equiv e^{-\gamma _{+}\left|t\right|}$ and $\phi _{-}(t)\equiv e^{-\gamma _{-}\left|t\right|}$ as basis functions $\bm{\phi }(t)$ (Regarding $\mathcal{S}(t)$, see Appendix \ref{sec:decomposition-brownian1}).
However, this basis set is degenerated in the critical--damped case, $\zeta =2\omega _{0}$, and this cannot express the non--exponential behavior of a critical--damped oscillator,
\begin{align}
  \mathcal{A}(t)&=-\lambda \omega _{0}^{2}\cdot te^{-(\zeta /2)\left|t\right|}.
\end{align}
Instead, we employ the basis
\begin{subequations}
  \begin{align}
    \phi _{p}(t)&=-\frac{\omega _{0}}{\omega _{1}}\sin (\omega _{1}\left|t\right|)e^{-\zeta \left|t\right|/2}\\
    \intertext{and}
    \phi _{q}(t)&=\left(\frac{\zeta }{2\omega _{1}}\sin (\omega _{1}\left|t\right|)+\cos (\omega _{1}\left|t\right|)\right)e^{-\zeta \left|t\right|/2},
  \end{align}
\end{subequations}
which satisfy $\phi _{p}(0)=0$, $\phi _{q}(0)=1$, and
\begin{align}
  \partial _{t}
  \begin{pmatrix}
    \phi _{p}(t)\\
    \phi _{q}(t)\\
  \end{pmatrix}
  &=-
  \begin{pmatrix}
    \zeta &\omega _{0}\\
    -\omega _{0}&0\\
  \end{pmatrix}
  \begin{pmatrix}
    \phi _{p}(t)\\
    \phi _{q}(t)\\
  \end{pmatrix}.
  \label{eq:brownian-basis}
\end{align}
Then $\mathcal{A}(t)$ is expressed as $\mathcal{A}(t)=\lambda \omega _{0}\phi _{p}(t)$.
When $\zeta =2\omega _{0}$, the coefficient matrix of Eq.~\eqref{eq:brownian-basis} is non--diagonalizable, while it is diagonalizable otherwise.
Note that, Eq.~\eqref{eq:brownian-basis} is equivalent to the phase space equations of motion of a damped harmonic oscillator with coordinate $q(t)$ and momentum $p(t)$,
\begin{subequations}
  \begin{align}
    p(t)&=-\zeta p(t)-\omega _{0}q(t)\\
    \intertext{and}
    q(t)&=+\omega _{0}p(t).
  \end{align}
\end{subequations}
This representation is related to phase--space Fokker--Planck equations \cite{tanimura2015jcp, ikeda2019jctc}.

\begin{figure}
  \centering
  \includegraphics[scale=\SingleColFigScale]{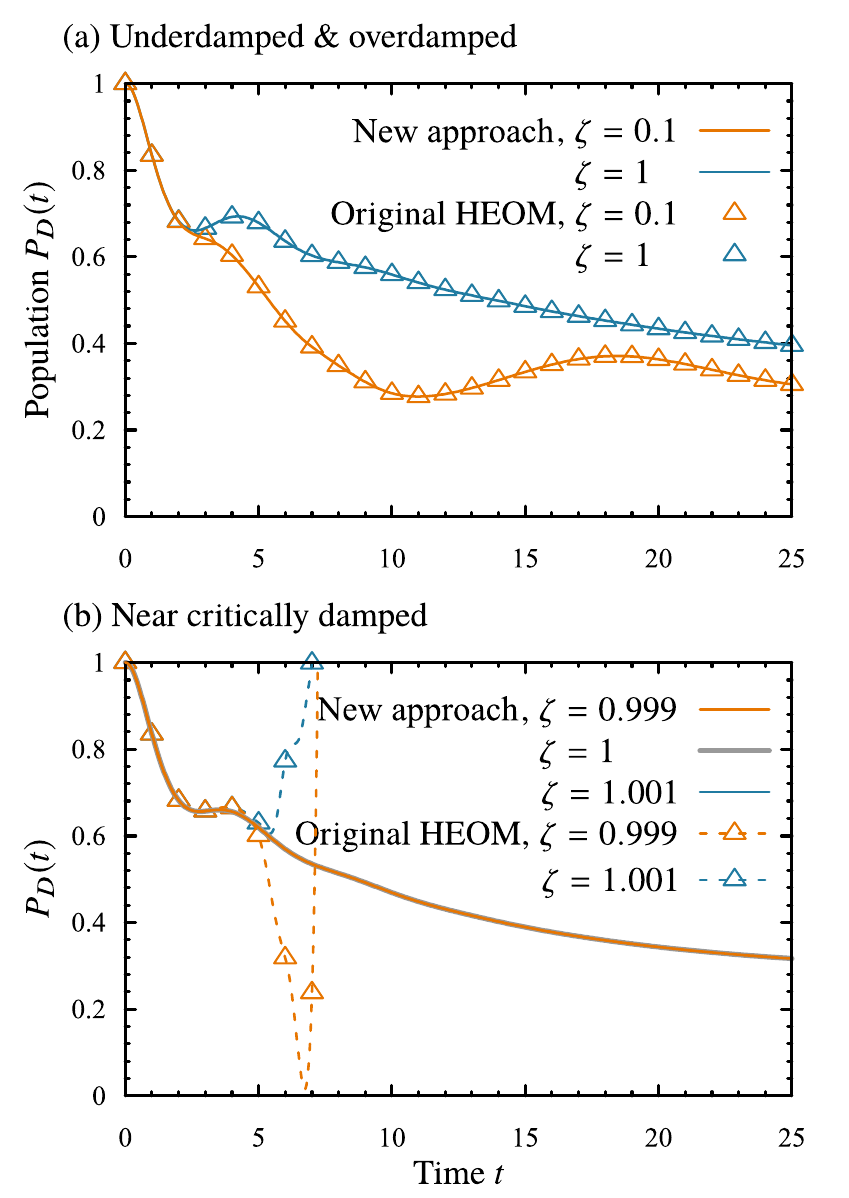}
  \caption{
    Donor population dynamics after excitation at $t=0$ in the case of (a) underdamped ($\zeta =0.1$), overdamped ($\zeta =2$), and (b) near critically damped cases $\zeta =0.999, 1, 1.001$.
    Orange, blue, and gray curves represent results of new treatment (Eq.~\eqref{eq:heom}), while orange and blue triangles represent results of the original HEOM (Eq.~\eqref{eq:oheom}).
    Here, $\lambda =2$, $\omega _{0}=0.5$, and $T=1$.
  }
  \label{fig:critically_damped}
\end{figure}
In Fig.~\ref{fig:critically_damped}, we show the numerical results in the cases of underdamped, critically damped, overdamped regions with our generalization Eq.~\eqref{eq:heom} and original HEOM Eq.~\eqref{eq:oheom}.
Here, we employed the Pad\`{e} spectral decomposition (PSD) $[N-1/N]$ scheme to express the Bose--Einstein distribution in $\mathcal{S}(t)$ \cite{hu2011jcp}, and adopted a single pole ($N=1$).
This causes an additional exponential function basis in $\mathcal{S}(t)$, and therefore $K=3$ including $\phi _{p}(t)$ and $\phi _{q}(t)$ (or $\phi _{+}(t)$ and $\phi _{-}(t)$).
The truncation tiers of the HEOM were chosen as $\mathcal{N}_{\mathrm{max}}=20$, $15$, and $25$ for $\lambda =0.01$, $0.1$ and $1$, respectively, which cause $1,770$, $815$, and $3,275$ ADOs.
For details, see Appendix \ref{sec:decomposition-brownian1}.

In the overdamped and underdamped regions, our generalization and original HEOM showed consistent results.
However, near critical--damped region, the original HEOM became very unstable, and was unsolvable with a finite $\mathcal{N}$.
This is because, the basis set used in the original HEOM, $\phi _{+}(t)$ and $\phi _{-}(t)$, are almost denegerate near the critically damped condition.
$\mathcal{A}(t)$ is described as a difference between $\phi _{+}(t)$ and $\phi _{-}(t)$, it causes numerical errors from loss of significance.
On the other hand, our generalization worked well in all regions because our method does not depend on diagonalization of the basis functions and therefore the stability is not related to degeneration.

As shown here, our new treatment can describe the dynamics caused by non--exponential correlation functions, which has a non--diagonalizable coefficient matrix $\bm{\gamma }$.
Moreover, even if the matrix $\bm{\gamma }$ is diagonalizable, the original framework can be unstable like the case of nearly critically damping regime.
Thus, our new treatment enables us to make more stable analysis for non--perturbative and non--Markovian regime.
Note that, in our treatment, the choice of basis set is not unique while the choice relates the stability of the equations.

\subsection{Spontaneous de--excitation under zero--temperature environment}
\label{sec:example3}
Finally, we discuss a spontaneous de--excitation process of a two--level system $\{|g\rangle , |e\rangle \}$ caused by zero--temperature bosonic environment.
Here, we assume that the Hamiltonian of the system and the system--bath interaction operator are expressed as $\hat{H}=\hbar \Omega _{e}\hat{a}^{+}\hat{a}^{-}$ and $\hat{V}=\hat{a}^{+}+\hat{a}^{-}$, respectively, where $\hat{a}^{+}=|e\rangle \langle g|$ and $\hat{a}^{-}\equiv |g\rangle \langle e|$ are creation/annihilation operators of the system.
As the environment model, we employ a Brownian spectral density
\begin{align}
  \mathcal{J}(\omega )&=2\alpha \frac{\zeta \omega _{0}^{2}\omega }{(\omega _{0}^{2}-\omega ^{2})^{2}+\zeta ^{2}\omega ^{2}},
  \label{eq:brownian-alpha}
\end{align}
which has a sharp peak at $\omega =\omega _{1}=\sqrt {\mathstrut \omega _{0}^{2}-\zeta ^{2}/4}$ in the underdamped condition $\zeta <2\omega _{0}$.
To make the system and bath characteristic frequencies resonant, we set $\Omega _{e}=\omega _{1}$.
Note that, although the coupling strength $\alpha $ has the same form as the reorganization energy $\lambda $ in Sec.~\ref{sec:example2}, physical meaning of the constants are quite different because of the difference of $\hat{V}$:
The interaction $\hat{V}$ here has no diagonal elements, and does not cause reorganization of the quantum states $|g\rangle $ and $|e\rangle $.

We set the initial state of the system as the pure excited state $\hat{\rho }(t_{0})=|e\rangle \langle e|$, and calculate population dynamics connected to zero--temperature bath $T=0$.
In this model, transitions among the two states are purely caused by fluctuation of the bath coordinates.
Because there is no thermal fluctuation in the zero--temperature limit, the spontaneous transition is caused by quantum fluctuations of the environment.

To apply the HEOM, we need to express $\mathcal{C}(t)$ in a basis set $\bm{\phi }(t)$.
It is difficult to treat $\mathcal{C}(t)$ in the case of low temperatures because we need to incorporate many bosonic Matsubara frequencies $2\pi k/\beta \hbar $ near a system frequency, which come from the Bose--Einstein distribution and provide quantum low--temperature corrections.
In the case of the zero--temperature, the bosonic Matsubara frequencies are almost degenerate and form a continuum.
There are many investigations to propose improved treatment for low--temperature situations and to eliminate unphysical artifacts caused by insufficient low--temperature corrections \cite{ishizaki2005jpsj, hu2011jcp, duan2017prb, dijkstra2017jcp, erpenbeck2018jcp, lambert2019nc, ishizaki2019jpsj}.
Here, we adopt the Fano spectral decomposition (FSD) technique recently made by Yan and co--workers \cite{cui2019jcp, zhang2020jcp}, which enables us to describe the low--temperature effects accurately with a few number of basis functions $\bm{\phi }(t)$.
This method divides the Bose--Einstein distribution into a high--temperature part with a reference temperature $T_{0}$ and a low--temperature part as $n_{\mathrm{BE}}(\omega ,T)=n_{\mathrm{BE}}(\omega ,T_{0})+\Delta n_{\mathrm{BE}}(\omega ,T,T_{0})$.
The high--temperature part is evaluated by using the conventional PSD framework, and the low--temperature part is accurately expressed by using summation of modified Fano functions.
The modified Fano functions result in basis functions in the forms of $t^{l}\cdot e^{-\gamma t}$ ($l\geq 0$), and $\bm{\gamma }$ caused by these functions are non--diagonalizable.
Hence, we cannot employ the original HEOM Eq.~\eqref{eq:oheom} and need generalizations for non--exponential functions.
In this article, we employ our new treatment Eq.~\eqref{eq:heom} to describe the dynamics under zero--temperature environment.
For details, see Appendix \ref{sec:decomposition-brownian2}.
Note that, it is impossible to perform the strict zero--temperature parametrization $T=0$ within the FSD framework.
However, if the temperature parameter we use is sufficiently small, $\mathcal{C}(t)$ asymptotically arrives the analytical solution of $T=0$ as discussed below.

To discuss the effects of the rigorous description based on our treatment, we introduce two approximation for comparison:

\paragraph{Rotating wave approximation}
In the zero--temperature limit, all of the bath modes are in their ground state when it is not connected to the system, and the factorized initial condition Eq.~\eqref{eq:fact-init} can be rewritten in the form of a wavefunction of the total system as
\begin{align}
  |\Psi ^{\mathrm{tot}}(t_{0})\rangle &=|e\rangle \otimes |\vec{0}\rangle .
  \label{eq:init-wf}
\end{align}
Here, $|\vec{0}\rangle $ represents the vaccum state of the bath.
By using creation and annihilation operators of the $\xi $th bath mode, $\hat{b}_{\xi }^{+}$ and $\hat{b}_{\xi }^{-}$, the interaction Hamiltonian of the total system can be rewritten as
\begin{align}
  \hat{H}^{\mathrm{int}}&=-\sum _{\xi }\frac{g_{\xi }}{\sqrt {\mathstrut 2}}\left(\hat{a}^{+}+\hat{a}^{-}\right)\left(\hat{b}_{\xi }^{+}+\hat{b}_{\xi }^{-}\right),
\end{align}
which includes resonant transitions $\hat{a}^{-}\hat{b}_{\xi }^{+}$ and $\hat{a}^{+}\hat{b}_{\xi }^{-}$ and non--resonant transitions $\hat{a}^{+}\hat{b}_{\xi }^{+}$ and $\hat{a}^{-}\hat{b}_{\xi }^{-}$.
By neglecting non--resonant terms, we obtain
\begin{align}
  \hat{H}^{\mathrm{int}}_{\mathrm{RWA}}&=-\sum _{\xi }\frac{g_{\xi }}{\sqrt {\mathstrut 2}}\left(\hat{a}^{-}\hat{b}_{\xi }^{+}+\hat{a}^{+}\hat{b}_{\xi }^{-}\right),
\end{align}
which is the well--known rotating wave approximation (RWA) form.

In this approximation, the number of excitations of the total wavefunction $|\Psi ^{\mathrm{tot}}(t)\rangle $ is preserved in the time evolution, and in the case of Eq.~\eqref{eq:init-wf}, the wavefunction is restricted in the single Fock state space as
\begin{align}
  |\Psi ^{\mathrm{tot}}(t)\rangle =c_{e,0}(t)|e\rangle \otimes |\vec{0}\rangle +\sum _{\xi }c_{g,\xi }(t)|g\rangle \otimes |\vec{1}_{\xi }\rangle .
  \label{eq:total-wavefunction}
\end{align}
Here, $|\vec{1}_{\xi }\rangle $ represents a state in which the $\xi $th mode is singly excited.
The dynamics of the coefficient $c_{e,0}(t)$ in the interaction picture,
\begin{align}
  \tilde{c}_{e,0}(t)&\equiv e^{+i\left(\hbar \omega _{0}+E_{\mathrm{vac}}\right)(t-t_{0})/\hbar }c_{e,0}(t),
\end{align}
is determined by a Volterra--type integro--differential equation (see Appendix \ref{sec:volterra})
\begin{align}
  \frac{\partial }{\partial t}\tilde{c}_{e,0}(t)&=-\frac{1}{\hbar ^{2}}\int _{t_{0}}^{t}\!ds\,\mathcal{C}(t-s)e^{+i\Omega _{e}(t-s)}\tilde{c}_{e,0}(s),
  \label{eq:volterra}
\end{align}
in which $\mathcal{C}(t)$ acts as a memory kernel.
Here, $E_{\mathrm{vac}}=\sum _{\xi }\hbar \omega _{\xi }/2$ is the summation of the zero--point energies of the bath.
This is equivalent to theories used in quantum electrodynamics \cite{dung2000pra, wang2019jcp}.
The population dynamics of the excited state can be calculated as $P_{e}(t)=\left|\tilde{c}_{0}(t)\right|^{2}$.

While it is possible to solve Eq.~\eqref{eq:volterra} directly, we can construct simpler time differential equations to solve the same problem as follows:
By substituting Eqs.~\eqref{eq:S-expansion} and \eqref{eq:A-expansion} into the Eq.~\eqref{eq:volterra},
\begin{widetext}
  \begin{align}
    \partial _{t}\tilde{c}_{e,0}(t)&=-\frac{1}{\hbar ^{2}}S_{\delta }\tilde{c}_{e,0}(s)-\frac{1}{\hbar ^{2}}\int _{t_{0}}^{t}\!ds\,(\vec{\sigma }^{t}(\bm{s}+i\bm{a})\vec{\phi }(t-s))e^{+i\omega _{0}(t-s)}\tilde{c}_{e,0}(s)\notag\\
    &=-\frac{1}{\hbar ^{2}}S_{\delta }\tilde{c}_{e,0}(t)-\frac{i}{\hbar }\sum _{k}\sigma _{k}\tilde{d}_{k}(t).
    \label{eq:diff-coeff-c}
  \end{align}
  Here, we have introduced auxiliary coefficients defined as
  \begin{align}
    \tilde{d}_{k}(t)&\equiv -\frac{i}{\hbar }\int _{t_{0}}^{t}\!ds\,\sum _{l}(s_{kl}+ia_{kl})\phi _{l}(t-s)e^{+i\Omega _{e}(t-s)}\tilde{c}_{e,0}(s).
  \end{align}
  The time evolution equation of $\tilde{d}_{j}(t)$ is evaluated as
  \begin{align}
    \partial _{t}\tilde{d}_{k}(t)&=+i\Omega _{e}\tilde{d}_{k}(t)-\sum _{l}\gamma _{kl}\tilde{d}_{k}(t)-\left[\sum _{l}\left(\frac{i}{\hbar }s_{kl}-\frac{1}{\hbar }a_{kl}\right)\phi _{l}(0)\right]\tilde{c}_{e,0}(t).
    \label{eq:diff-coeff-d}
  \end{align}
\end{widetext}
Hence, the set of equations Eqs.~\eqref{eq:diff-coeff-c} and \eqref{eq:diff-coeff-d} are equivalent to Eq.~\eqref{eq:volterra}.
This can be regarded a generalization of a method to solve a Volterra--type integro--differential equation by using a second--order differential equation in Ref \onlinecite{wang2019jcp}, and a similar method is found in Ref.~\onlinecite{elattari2000pra} in the case of a multi--exponential function basis set.
Remarkably, this set of equations has the same structure to our HEOM Eq.~\eqref{eq:heom-int} truncated by the first--tier $\mathcal{N}\leq 1$.
Because of the RWA, the hierarchical elements which are higher--order tiers are not occupied.

\paragraph{Single Lorentzian component approximation}
In the case of underdamped condition $\omega _{0}>\zeta /2$, the Brownian spectral density can be decomposed into two Lorentzian components as
\begin{align}
  \mathcal{J}(\omega )&=\frac{\alpha }{2\omega _{1}}\left(\frac{\gamma _{\mathrm{c}}\omega _{0}^{2}}{(\omega -\omega _{1})^{2}+(\gamma _{\mathrm{c}}/2)^{2}}-\frac{\gamma _{\mathrm{c}}\omega _{0}^{2}}{(\omega +\omega _{1})^{2}+(\gamma _{\mathrm{c}}/2)^{2}}\right)\notag\\
  &\equiv \mathcal{J}_{+}(\omega )+\mathcal{J}_{-}(\omega ),
\end{align}
where $\omega _{1}\equiv \sqrt {\mathstrut \omega _{0}^{2}-\gamma _{\mathrm{c}}^{2}/2}$ and $\mathcal{J}_{\pm }(\omega )$ are resonant to $\omega =\pm \omega _{1}$.
When the Lorenzian components are sufficiently narrow, i.e.~$\gamma _{\mathrm{c}}$ is sufficiently small in comparison with $\omega _{1}$, we can neglect $\mathcal{J}_{-}(\omega )$ in the region $\omega >0$ and we can regard $\mathcal{J}_{+}(\omega )\simeq 0$ in the region $\omega <0$.
Under this approximation, the quantum correlation function $\mathcal{C}(t)$ in the zero--temperature is evaluated as
\begin{align}
  \mathcal{C}(t)&\simeq \mathcal{C}_{+}(t)\equiv \frac{1}{\pi }\int _{-\infty }^{\infty }\!d\omega \,\mathcal{J}_{+}(\omega )e^{-i\omega t}.
  \label{eq:single-lorentz}
\end{align}
Here, we have used $n_{\mathrm{BE}}(\omega ,+0)+1=\theta _{\mathrm{H}}(\omega )$, where $\theta _{\mathrm{H}}(x)$ is the Heaviside step function.
Note that while this approximate form of spectral density, $\mathcal{J}_{+}(\omega )$, is sometimes adopted in open quantum theories, the spectral density does not satisfy $\mathcal{J}_{+}(-\omega )=-\mathcal{J}_{+}(\omega )$.

Hereafter, we discuss the effects of rigorous description (New approach, Eq.~\eqref{eq:heom}), wavefunction description with the rotating wave approximation (WF--RWA, Eqs.~\eqref{eq:diff-coeff-c} and \eqref{eq:diff-coeff-d}), and Born--Markov approximation (Redfield equation).
  

\begin{figure}
  \centering
  \includegraphics[scale=\SingleColFigScale]{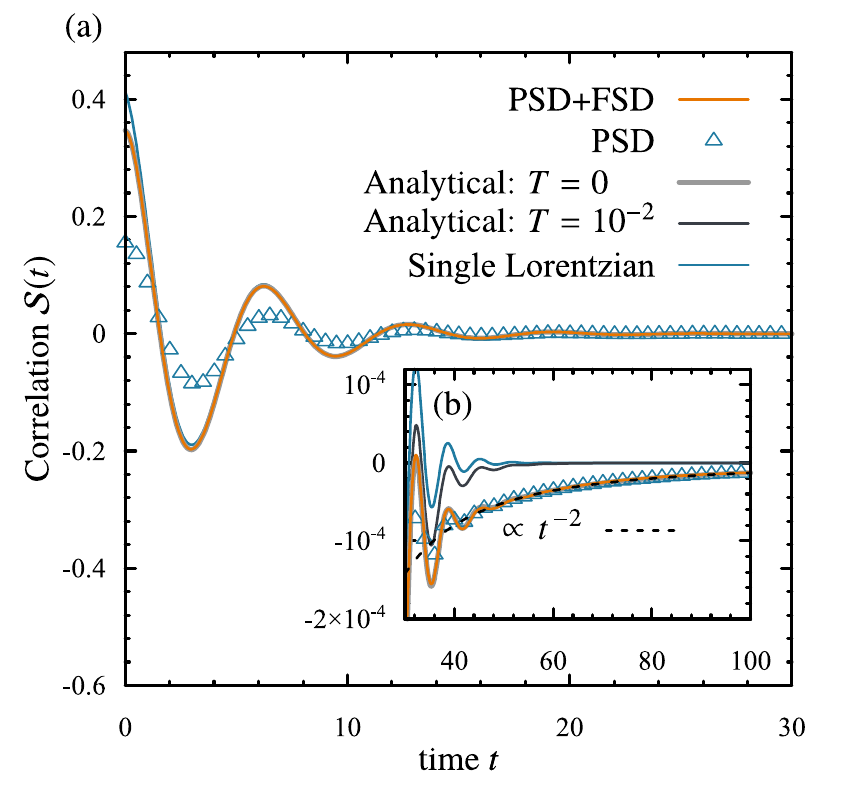}
  \caption{ (a) Symmetrized correlation function $\mathcal{S}(t)$ in the case of zero--temperature limit with a Brownian spectral density Eq.~\eqref{eq:brownian-alpha}.
    Orange curves and blue triangles represent the results of the PSD$[0/1]$ +FSD$[9]$ scheme and PSD$[9/10]$ scheme for $T=10^{-3}$, respectively.
    These two cause $12$ basis functions $\bm{\phi }(t)$.
    Blue curves mean the results of single Lorentzian approximation Eq.~\eqref{eq:single-lorentz}, and gray and black curves represent analytical solutions with strict zero--temperature $T=0$ and relatively high--temperature $T=10^{-2}$ cases, respectively.
    (b) Long--time tail behavior of (a).
  }
  \label{fig:zerotemp_correlation}
\end{figure}
To check the accuracy of correlation function expansions, we first plotted the symmetrized correlation function $\mathcal{S}(t)$ in Fig.~\ref{fig:zerotemp_correlation}.
Here, we employ the FSD scheme with temperature $T=10^{-3}$ and a single PSD$[N-1/N]$ pole on the reference temperature $T_{0}=1$ (PSD$[0/1]$+FSD$[9]$), which yields $12$ basis functions $\bm{\phi }(t)$.
As seen here, the results of this expansion (orange curves) sufficiently reproduces the analytical solution of the strict zero--temperature case (gray curves).
The analytical solution and expansion has an algebraic long--time tail with power $t^{-2}$, which is a characteristic feature of a Brownian motion near zero--temperature under an Ohmic friction \cite{jung1985pra, weiss2011book}.
Although the single Lorentzian approximation Eq.~\eqref{eq:single-lorentz} also well reproduces the analytical solution, the amplitude near $t=0$ is different from the correct solution, and this cannot reproduce the long--time tail behavior.

\begin{figure}
  \centering
  \includegraphics[scale=\SingleColFigScale]{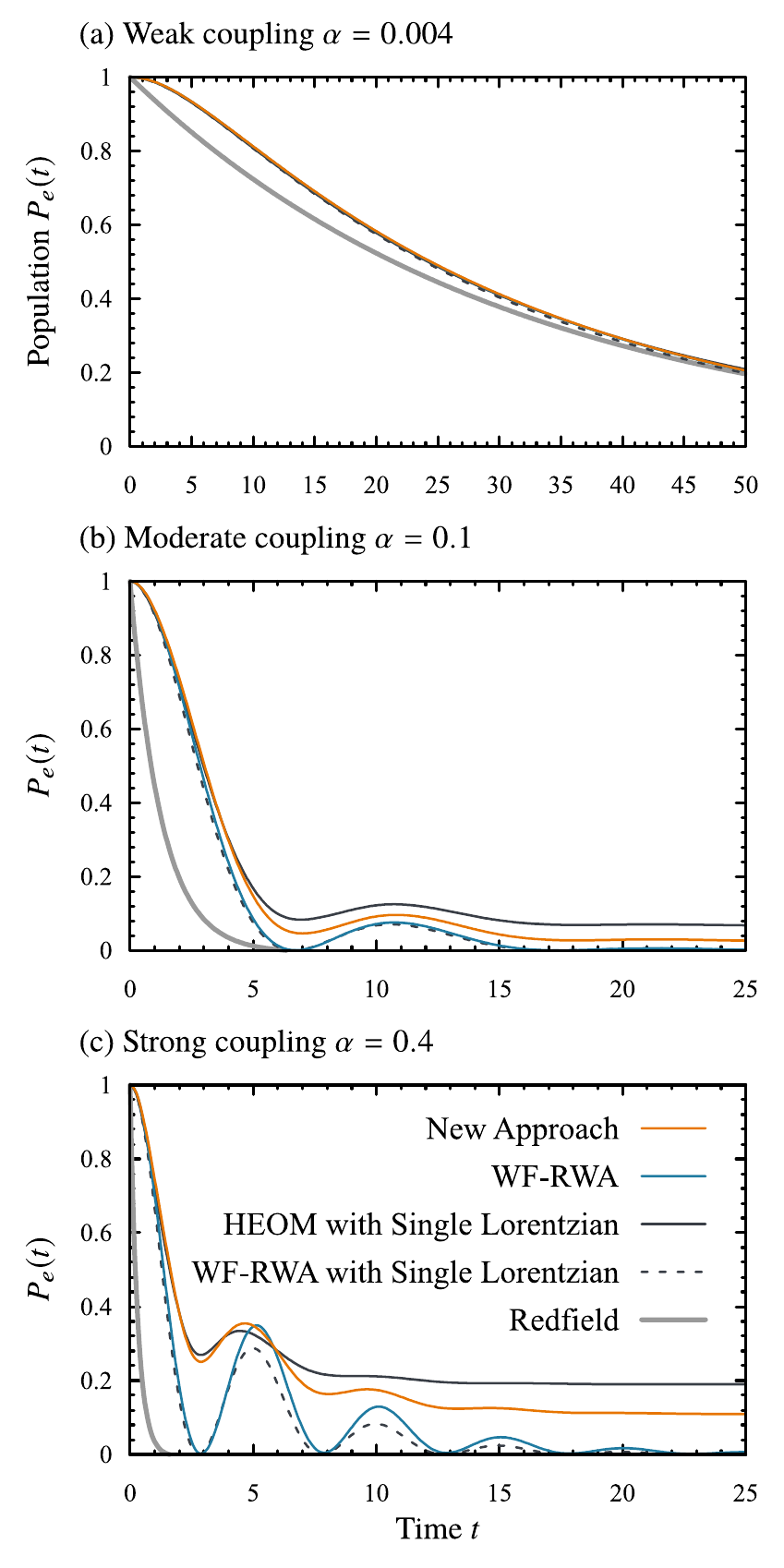}
  \caption{ Spontaneous de--excitation processes of the excited population $P_{e}(t)\equiv \rho _{ee}(t)$ under a zero--temperature environment for (a) weak coupling $\alpha =0.004$, (b) moderate coupling $\alpha =0.1$, and (c) strong coupling $\alpha =0.4$ cases.
    Orange curves represent our generalization of the HEOM Eq.~\eqref{eq:heom} with the PSD$[0/1]$+FSD$[9]$ parametrization, and blue curves mean the results of the WF--RWA calculations, Eqs.~\eqref{eq:diff-coeff-c} and \eqref{eq:diff-coeff-d}, with the same basis set $\bm{\phi }(t)$.
    Black solid and dashed curves represent the results of the HEOM and WF--RWA calculations with single Lorentzian approximation Eq.~\eqref{eq:single-lorentz}.
    Gray curves mean the results of the Redfield theory.
  }
  \label{fig:zerotemp_population}
\end{figure}
In Fig.~\eqref{fig:zerotemp_population}, we depict the numerical results of the spontaneous de--excitation of the excited state population in the case of weak, moderate, and strong coupling cases.
The parameters were set to $\omega _{0}=1$ and $\zeta =0.5$, and this causes $\omega _{1}=\Omega _{e}=0.968$.
The truncation tiers of the HEOM were chosen as $\mathcal{N}_{\mathrm{max}}=3$, $5$, and $8$ for $\alpha =0.004$, $0.1$ and $0.4$, respectively, which generates $454$, $6,187$, and $125,969$ ADOs.

In Fig.~\ref{fig:zerotemp_population}(a), the results of the Redfield theory and other calculations are already different even though it is in a weak coupling regime.
This is because, the non--Markovian feature of the environment correlation is important in this problem:
In the cases of Sections~\ref{sec:example1} and \ref{sec:example2}, the population transfer processes are caused by the electronic coupling of the system states, $J$, and the beating behavior in the weak coupling regime is dominantly determined by the coupling.
On the other hand, in the problem of this section, the transitions are caused by the bath fluctuations and their behavior reflect details of the bath correlation functions in comparison with the previous problems.
Because both HEOM and WF--RWA calculations capture the non--Markovian behavior, the results are coincident.

In the moderate and strong coupling cases, Figs.~\ref{fig:zerotemp_population}(b) and (c), the WF--RWA results differ from the HEOM results.
This is because, when the system--bath coupling is strong, the non--resonant terms $\hat{a}^{+}\hat{b}_{\xi }^{+}$ and $\hat{a}^{-}\hat{b}_{\xi }^{-}$ which are neglected in the WF--RWA calculations contribute de--excitation processes.
In this regime, the excited state has a finite equilibrium population caused by quantum fluctuations of the bath even though it is zero--temperature (In the strong--coupling limit, the population should be $0.5$ because the enegy difference between ground and excited states becomes negligible).
The WF--RWA calculations underestimate the population owing to the lack of non--resonant terms.
The results of single Lorentzian approximation become different from the correct calculations when the coupling is strong, because details of the correlation functions affect in the regime.

\section{CONCLUDING REMARKS}
\label{sec:conclusion}
In this paper, we developed a new generalization of the HEOM theory including treatments of non--exponential basis sets for environment correlation functions, Eq.~\eqref{eq:heom}.
We showed that our generalization was unified with the original HEOM theory and other generalizations, and our generalization could be more efficient and stable than the conventional theories.
We demonstrated our new generalization by using three examples, in which non--exponential behavior of environment correlation functions plays essential roles, and we further examined the validity of our approach.

Our generalization is based on a new, simple design of expansion forms of environment correlation functions, Eqs.~\eqref{eq:S-expansion} and \eqref{eq:A-expansion}, and while other elements of the theory are almost the same as the original HEOM theory.
Therefore, other techniques developed within the framework of the original theory, e.g.,~the HEOM theory for fermion environments \cite{jin2007jcp, jin2008jcp, hartle2013prb, hartle2015prb, schinabeck2018prb}, should be easily incorporated with our generalization.

In our demonstration calculations, we employed a simple truncation of tiers of the HEOM, which could be inefficient.
An extension of more advanced truncation scheme for our generalization, e.g., those in Refs.~\onlinecite{hartle2013prb, hartle2015prb}, should be worthful for more efficient, stable, and general calculations.
In addition, in this paper, we discussed only the instability of HEOM in the case of near degenerate basis functions.
Besides this, instabilities of HEOM theories in the case of strongly non--Markovian bath correlation function are known.
Although a numerical approach to remove the instabilities has been proposed\cite{dunn2019jcp}, the physics underlying the instabilities is an unsolved problem.
Improving the stability of HEOM will be worthful for many practical applications.
These problems are left for future investigations.

\section*{SUPPLEMENTARY MATERIAL}
See supplementary material for the LibHEOM library and its Python binding (PyHEOM) which we developed to perform simulations based on our approach. 
Their documents and up--to--date codes may be found on GitHub.

\begin{acknowledgments}
  T.~Ikeda is thankful to B.~Fu for manuscript reading.
  T.~Ikeda was supported by JSPS Overseas Challenge Program for Young Researchers.
  G.~D.~Scholes is a CIFAR Fellow in the Bio--Inspired Energy Program.
  This research is funded by the Gordon and Betty Moore Foundation through Grant GBMF7114.
\end{acknowledgments}

\section*{DATA AVAILABILITY}
Data available on request from the authors.

\appendix
\section{Construction of matrices \texorpdfstring{$\bm{s}$}{s} and \texorpdfstring{$\bm{a}$}{a} commuting with \texorpdfstring{$\bm{\gamma }$}{gamma} }
\label{sec:commuting-s-a}
In this section, we give examples of possible constructions of $\bm{\sigma }$, $\bm{s}$, and $\bm{a}$ form $\bm{S}$ and $\bm{A}$.
Hereafter, we assume that the vector $\bm{\sigma }$ is properly given.
The choice $\sigma _{k}=1$ is sufficient, while other choises are possible.
Hereafter, $\bm{b}$ represents $\bm{s}$ or $\bm{a}$, and $\bm{B}$ is the corresponding $\bm{S}$ or $\bm{A}$.

\subsection{Diagonalizable \texorpdfstring{$\bm{\gamma }$}{gamma}}
\label{sec:commuting-s-a-1}
When $\bm{\gamma }$ is diagonalizable, $\bm{b}$ can be constructed as a simultaneously diagonalizable matrix.
We assume that $\bm{\gamma }$ is diagonalized as $\bm{Z}^{-1}\bm{\gamma }\bm{Z}=\bm{\gamma }'$, where $\bm{\gamma }'$ a diagonal matrix.
From Eq.~\eqref{eq:S-s-relation}, $\bm{b}$ is constructed as $\bm{b}=\bm{Z}\bm{b}'\bm{Z}^{-1}$, where $\bm{b}'$ is a diagonal matrix which diagonal elements are given by $b'_{k,k}=({}^{t}\bm{B}\bm{Z})_{k}/({}^{t}\bm{\sigma }\bm{Z})_{k}$.

\subsection{Non--diagonalizable \texorpdfstring{$\bm{\gamma }$}{gamma}}
\label{sec:commuting-s-a-2}
When $\bm{\gamma }$ is non--diagonalizable, it is possible to construct $\bm{b}$ by expressing it as a linear combination of $\bm{\gamma }^{k}$ ($k=0,\dots ,K-1$), i.e., $\bm{b}=\sum _{k=0}^{K-1}\tilde{b}_{k}\bm{\gamma }^{k}$.
From Eq.~\eqref{eq:S-s-relation}, the coefficients $\tilde{b}_{k}$ should satisfy $\bm{B}=\sum _{k=0}^{K-1}\tilde{b}_{k}\bm{u}_{k}$, where $\bm{u}_{k}\equiv {}^{t}\bm{\gamma }^{k}\bm{\sigma }$.
Therefore, the coefficients $\tilde{b}_{k}$ can be constructed by orthogonalizing a ``basis set'' $\bm{u}_{k}$ and by expressing $\bm{B}$ by the orthogonalized basis set.
The orthogonalization is, e.g., performed by calculation of an inversed matrix or the Gram--Schmidt process, and in the case of former, the coefficients are given $\tilde{b}_{k}=(\bm{U}^{-1}\bm{B})_{k}$ where $\bm{U}$ is a matrix defined as $\bm{U}\equiv (\bm{u}_{0}~\dots ~\bm{u}_{K-1})$.
Because this method causes errors from loss of significance in the case of a large matrix $\bm{\gamma }$, it is more proper to decompose $\bm{\gamma }$ into small block matrices $\bm{\gamma }_{k}^{\mathrm{Block}}$ and perform evaluations of $\bm{b}_{k}^{\mathrm{Block}}$ by using this method.

\section{Relation to Other Generalizations of HEOM}
\label{eq:rel-yheom}
First, we show the relation between our new approach and a generalization of the HEOM theory given in Ref.~\onlinecite{xu2005jcp}.
Similar to the case of the extended HEOM in Sec.~\ref{sec:rel-eheom}, we duplicate the basis as $^{t}\bm{\phi }'(t)=(^{t}\bm{\phi }(t),^{t}\bm{\phi }(t))$, which satisfies Eq.~\eqref{eq:gamma_prime}.
When we choose the expansion of correlation functions $\mathcal{S}(t)$ and $\mathcal{A}(t)$ as
\begin{subequations}
  \begin{align}
    &\begin{aligned}
       \mathcal{S}(t)&=
       {}^{t}\bm{\sigma }'\bm{s}'\bm{\phi }'(t)\equiv 
       \begin{pmatrix}
         -i{}^{t}\bm{\sigma }_{\mathcal{C}}&i{}^{t}\bm{\sigma }_{\mathcal{C}}^{\star }\\
       \end{pmatrix}
       \begin{pmatrix}
         -\bm{1/2i}&\bm{0}\\
         \bm{0}&+\bm{1/2i}\\
       \end{pmatrix}
       \begin{pmatrix}
         \bm{\phi }(t)\\
         \bm{\phi }(t)\\
       \end{pmatrix}\\
       &=\sum _{k}\frac{\sigma _{\mathcal{C},k}+\sigma _{\mathcal{C},k}^{\star }}{2}\phi _{k}(t)
     \end{aligned}
    \intertext{and}
    &\begin{aligned}
       \mathcal{A}(t)&=
       {}^{t}\bm{\sigma }'\bm{a}'\bm{\phi }'(t)\equiv 
      \begin{pmatrix}
        -i{}^{t}\bm{\sigma }_{\mathcal{C}}&i{}^{t}\bm{\sigma }_{\mathcal{C}}^{\star }\\
      \end{pmatrix}
      \begin{pmatrix}
        \bm{1/2}&\bm{0}\\
        \bm{0}&\bm{1/2}\\
      \end{pmatrix}
      \begin{pmatrix}
        \bm{\phi }(t)\\
        \bm{\phi }(t)\\
      \end{pmatrix}\\
      &=\sum _{k}\frac{\sigma _{\mathcal{C},k}-\sigma _{\mathcal{C},k}^{\star }}{2i}\phi _{k}(t),
    \end{aligned}
  \end{align}
\end{subequations}
where $\sigma _{\mathcal{C},k}\equiv S_{k}+iA_{k}$ and $\sigma _{\mathcal{C},k}^{\star }\equiv S_{k}-iA_{k}$, we obtain the corresponding HEOM from Eq.~\eqref{eq:heom} as
\begin{widetext}
  \begin{align}
    \begin{split}
      \partial _{t}\hat{\rho }_{\bm{l},\bm{m}}(t)&=-\mathcal{L}\hat{\rho }_{\bm{l},\bm{m}}(t)\\
      &\quad -\sum _{j,k}\gamma _{jk}l_{j}\hat{\rho }_{\bm{l}-\bm{1}_{j}+\bm{1}_{k},\bm{m}}(t)-\sum _{j,k}\gamma _{jk}m_{j}\hat{\rho }_{\bm{l},\bm{m}-\bm{1}_{j}+\bm{1}_{k}}(t)\\
      &\quad +\sum _{k}\hat{\Phi }\left(i\sigma _{\mathcal{C},k}\hat{\rho }_{\bm{l}+\bm{1}_{k},\bm{m}}(t)-i\sigma _{\mathcal{C},k}^{\star }\hat{\rho }_{\bm{l},\bm{m}+\bm{1}_{k}}(t)\right)\\
      &\quad +\sum _{k}\phi _{k}(0)l_{k}(\hat{V}^{\rightarrow }/\hbar )\hat{\rho }_{\bm{l}-\bm{1}_{k},\bm{m}}(t)+\sum _{k}\phi _{k}(0)m_{k}(\hat{V}^{\leftarrow }/\hbar )\hat{\rho }_{\bm{l},\bm{m}-\bm{1}_{k}}(t).
      \label{eq:heom-yan}
    \end{split}
  \end{align}
  Here, the ADOs have been defined as
  \begin{align}
    \begin{split}
      \tilde{\rho }_{\bm{l},\bm{m}}(t)&\equiv \mathcal{T}_{+}\prod _{k}\left(\int _{t_{0}}^{t}\!ds\,\phi _{k}(t-s)\tilde{V}^{\rightarrow }(s)/\hbar \right)^{l_{k}}\prod _{k}\left(\int _{t_{0}}^{t}\!ds\,\phi _{k}(t-s)\tilde{V}^{\leftarrow }(s)/\hbar \right)^{m_{k}}
      \mathcal{F}(t,t_{0})\tilde{\rho }(t_{0}).
    \end{split}
  \end{align}
\end{widetext}
The matrices $\bm{s}'$ and $\bm{a}'$ commute with $\bm{\gamma }'$, and $\sigma _{\mathcal{C},k}^{\star }=\sigma _{\mathcal{C},k}^{\ast }$ when $S_{k}$ and $A_{k}$ are real numbers.
When the basis set is given as $^{t}\bm{\phi }(t)=(te^{-\gamma t}, e^{-\gamma t})$, Eq.~\eqref{eq:heom-yan} reduces to an example of HEOM given in Ref.~\onlinecite{xu2005jcp}.
Note that Eq.~\eqref{eq:heom-yan} has the same hierarchical structure as the extended HEOM.
Thus, our new treatment can be regarded as an efficient generalization of the HEOM in Ref.~\onlinecite{xu2005jcp}, because our treatment does not require doubling the number of indices of the hierarchy.

Next, we consider a situation in which non--exponential basis functions appear only in the symmetrized correlation function.
The symmetrized and anti--symmetrized correlation functions can be expressed as $\mathcal{S}(t)=\sum _{k}S_{k}^{\text{exp}}\phi _{k}^{\text{exp}}(t)+\sum _{k'}S_{k'}^{\text{non--exp}}\phi _{k'}^{\text{non--exp}}(t)+S_{\delta }\cdot 2\delta (t)$ and $\mathcal{A}(t)=\sum _{k}A_{k}^{\text{exp}}\phi _{k}^{\text{exp}}(t)$.
Here, $\bm{\phi }^{\text{exp}}(t)$ and $\bm{\phi }^{\text{non--exp}}(t)$ are sets of exponential and non--exponential basis functions, respectively.
We assume that the basis set $^{t}\bm{\phi }(t)=(^{t}\bm{\phi }^{\text{exp}}(t),^{t}\bm{\phi }^{\text{non--exp}}(t))$ satisfies
\begin{align}
  \partial _{t}
  \begin{pmatrix}
    \bm{\phi }^{\text{exp}}(t)\\
    \bm{\phi }^{\text{non--exp}}(t)\\
  \end{pmatrix}
  &=
  -
  \begin{pmatrix}
    \bm{\gamma }^{\text{exp}} & \bm{0}\\
    \bm{0} & \bm{\gamma }^{\text{non--exp}}\\
  \end{pmatrix}
  \begin{pmatrix}
    \bm{\phi }^{\text{exp}}(t)\\
    \bm{\phi }^{\text{non--exp}}(t)\\
  \end{pmatrix},
  \label{eq:exp-non-exp-decomp}
\end{align}
where $\bm{\gamma }^{\text{exp}}$ is a diagonal matrix, $\{\bm{\gamma }^{\text{exp}}\}_{kk}=\gamma _{k}^{\text{exp}}$.
In this case, the correlation functions can be decomposed as
\begin{widetext}
  \begin{subequations}
    \begin{align}
      &\begin{aligned}
         \mathcal{S}(t)&=
         {}^{t}\bm{\sigma }\bm{s}\bm{\phi }(t)\equiv 
         \begin{pmatrix}
           ^{t}\bm{1}&^{t}\bm{S}^{\text{non--exp}}\\
         \end{pmatrix}
         \begin{pmatrix}
           \bm{s}^{\text{exp}}&\bm{0}\\
           \bm{0}&\bm{0}\\
         \end{pmatrix}
         \begin{pmatrix}
           \bm{\phi }^{\text{exp}}(t)\\
           \bm{\phi }^{\text{non--exp}}(t)\\
         \end{pmatrix}
         +S_{\delta }\cdot 2\delta (t)
       \end{aligned}
      \intertext{and}
      &\begin{aligned}
         \mathcal{A}(t)&=
         {}^{t}\bm{\sigma }\bm{a}\bm{\phi }(t)\equiv 
         \begin{pmatrix}
           ^{t}\bm{1}&^{t}\bm{S}^{\text{non--exp}}\\
         \end{pmatrix}
         \begin{pmatrix}
           \bm{a}^{\text{exp}}&\bm{0}\\
           \bm{0}&\bm{1}\\
         \end{pmatrix}
         \begin{pmatrix}
           \bm{\phi }^{\text{exp}}(t)\\
           \bm{\phi }^{\text{non--exp}}(t)\\
         \end{pmatrix},
       \end{aligned}
    \end{align}
  \end{subequations}
\end{widetext}
where $\bm{s}^{\text{exp }}$ and $\bm{a}^{\text{exp }}$ have only diagonal elements, $\{\bm{s}^{\text{exp }}\}_{kk}=S_{k}^{\text{exp}}$ and $\{\bm{a}^{\text{exp }}\}_{kk}=a_{k}^{\text{exp}}$, and $\bm{s}$ and $\bm{a}$ commute with $\bm{\gamma }$.
Then the corresponding HEOM can be expressed as
\begin{widetext}
  \begin{align}
    \begin{split}
      \partial _{t}\hat{\rho }_{\bm{n},\bm{n}'}(t)&=-(\mathcal{L}+\hat{\Xi })\hat{\rho }_{\bm{n},\bm{n}'}(t)-\sum _{k}n_{k}\gamma _{k}^{\text{exp}}\hat{\rho }_{\bm{n}}(t)-\sum _{j',k'}n_{j'}'\gamma _{j'k'}^{\text{non--exp}}\hat{\rho }_{\bm{n},\bm{n}'-\bm{1}_{j'}+\bm{1}_{k'}}(t)\\
      &\quad -\sum _{k}\hat{\Phi }\hat{\rho }_{\bm{n}+\bm{1}_{k},\bm{n}'}(t)-\sum _{k}n_{k}(S_{k}^{\text{exp}}\phi _{k}^{\text{exp}}(0)\hat{\Phi }-A_{k}^{\text{exp}}\phi _{k}^{\text{exp}}(0)\hat{\Psi })\hat{\rho }_{\bm{n}-\bm{1}_{k},\bm{n}'}(t)\\
      &\quad -\sum _{k'}S_{k'}^{\text{non--exp}}\hat{\Phi }\hat{\rho }_{\bm{n},\bm{n'}+\bm{1}_{k'}}(t)-\sum _{k'}n'_{k'}\phi _{k'}^{\text{non-exp}}(0)\hat{\Phi }\hat{\rho }_{\bm{n},\bm{n'}-\bm{1}_{k'}}(t).
      \label{eq:yheom2}
    \end{split}
  \end{align}
\end{widetext}
Here $\bm{n}$ and $\bm{n}'$ represents hierarchy indices for exponential and non--exponential basis functions, respectively.
This HEOM can be regarded as a combination between the original HEOM and extended HEOM, while, the number of indices of hierarchy equals to the number of basis functions.
Note that a polynomial--exponential function basis set does not satisfy Eq.~\eqref{eq:exp-non-exp-decomp}.

\section{Coefficients of \texorpdfstring{$\mathcal{S}(t)$}{S(t)} and \texorpdfstring{$\mathcal{A}(t)$}{A(t)} in Sec.~\ref{sec:example1}}
\label{sec:decomposition-superohmic}
By inserting Eq.~\eqref{eq:semicircle} into Eq.~\eqref{eq:C}, the integral reduces to a definite integral with interval $-\gamma _{\mathrm{c}}\leq \omega \leq \gamma _{\mathrm{c}}$.
By changing the variable of integration from $\omega $ to $\theta $ with the relation $\omega =\gamma _{\mathrm{c}}\sin \theta $, and by using the Jacobi--Anger expansion
\begin{subequations}
  \begin{align}
    \cos (x\sin \theta )&=J_{0}(x)+2\sum _{n=1}^{\infty }J_{2n}(x)\cos [2n\theta ]\\
    \intertext{and}
    \sin (x\sin \theta )&=2\sum _{n=1}^{\infty }J_{2n-1}(x)\sin [(2n-1)\theta ],
  \end{align}
\end{subequations}
we get the coefficients in Eqs.~\eqref{eq:S-bessel} and \eqref{eq:A-bessel} as
\begin{subequations}
  \begin{align}
    &\left\{
    \begin{aligned}
      &S_{0}=\frac{2\lambda (1+\sum _{j}2\eta _{j})}{\beta \hbar }-\sum _{j}\frac{4\lambda \eta _{j}\nu _{j}}{\beta \hbar }\frac{B_{j}^{2}}{R_{j}}X_{j},\\
      &S_{2}=-\sum _{j}\frac{4\lambda \eta _{j}\nu _{j}}{\beta \hbar }\frac{B_{j}^{4}}{R_{j}}X_{j},\\
      &S_{4}=-\frac{2\lambda (1+\sum _{j}2\eta _{j})}{\beta \hbar }+\sum _{j}\frac{4\lambda \eta _{j}\nu _{j}}{\beta \hbar }\frac{B_{j}^{4}}{R_{j}}X_{j}^{2},\\
      &S_{2k}=\sum _{j}\frac{4\lambda \eta _{j}\nu _{j}}{\beta \hbar }\frac{B_{j}^{2k}}{R_{j}}X_{j}^{2},\quad (k\geq 3),\\
      &S_{2k+1}=0\quad (k\geq 0)
    \end{aligned}
    \right.
    \intertext{and}
    &\left\{
    \begin{gathered}
      A_{1}=-\lambda \gamma _{\mathrm{c}},\quad A_{3}=-\frac{\lambda \gamma _{\mathrm{c}}}{2},\quad A_{5}=+\frac{\lambda \gamma _{\mathrm{c}}}{2},\\
      A_{2k}=0~(k\geq 0),
    \end{gathered}
    \right.
  \end{align}
\end{subequations}
respectively.
Here,
\begin{align}
  B_{j}&\equiv \frac{\gamma _{\mathrm{c}}}{R_{j}+\nu _{j}},&&&X_{j}&\equiv B_{j}^{-2}-B_{j}^{2},\notag\\
  \intertext{and}
  R_{j}&\equiv \sqrt {\gamma _{\mathrm{c}}\mathstrut ^{2}+\nu _{j}^{2}},
\end{align}
and we have introduced an expansion of $n_{\mathrm{BE}}(\omega )$ as
\begin{align}
  n_{\mathrm{BE}}(\omega )+\frac{1}{2}&\simeq \frac{1}{\beta \hbar \omega }+\sum _{j}\frac{2\eta _{j}}{\beta \hbar }\frac{\omega }{\omega ^{2}+\nu _{j}^{2}}.
  \label{eq:n-decomposition}
\end{align}
For an infinite number of $j$, the expansion coefficients should be $\eta _{j}=1$ and $\nu _{j}=2\pi j/\beta \hbar $, and $\nu _{j}$ is the $j$th bosonic Matsubara frequency.
When we want to increase the efficiency of the summation of $j$ in a finite number, the PSD$[N{-}1/N]$ scheme \cite{hu2011jcp} should be better choice.

\section{Coefficients of \texorpdfstring{$\mathcal{S}(t)$}{S(t)} and \texorpdfstring{$\mathcal{A}(t)$}{A(t)} in Sec.~\ref{sec:example2}}
\label{sec:decomposition-brownian1}
By inserting Eqs.~\eqref{eq:brownian} and \eqref{eq:n-decomposition} into Eqs.~\eqref{eq:S} and \eqref{eq:A} and by using the residue theorem, we obtain
\begin{subequations}
  \begin{align}
    \begin{split}
      \mathcal{S}(t)&=-\frac{i\lambda \omega _{0}^{2}}{\beta \hbar \omega _{1}}\left(\frac{1}{\gamma _{+}^{2}}+2\sum _{k=1}^{K}\frac{\eta _{k}}{\gamma _{+}^{2}-\nu _{k}^{2}}\right)\gamma _{+}e^{-\gamma _{+}\left|t\right|}\\
      &\quad +\frac{i\lambda \omega _{0}^{2}}{\beta \hbar \omega _{1}}\left(\frac{1}{\gamma _{-}^{2}}+2\sum _{k=1}^{K}\frac{\eta _{k}}{\gamma _{-}^{2}-\nu _{k}^{2}}\right)\gamma _{-}e^{-\gamma _{-}\left|t\right|}\\
      &\quad +\sum _{k}^{K}2\eta _{k}\frac{2\lambda }{\beta \hbar }\left(\frac{\nu _{k}\omega _{0}^{2}\zeta }{(\nu _{\kappa }^{2}+\omega _{0}^{2})^{2}-\zeta ^{2}\nu _{\kappa }^{2}}\right)e^{-\nu _{k}\left|t\right|}
    \end{split}
  \end{align}
  and 
  \begin{align}
    \mathcal{A}(t)&=\frac{\lambda \omega _{0}^{2}i}{2\omega _{1}}\left(e^{-\gamma _{+}\left|t\right|}-e^{-\gamma _{-}\left|t\right|}\right).
  \end{align}
\end{subequations}
By using relation
\begin{subequations}
  \begin{align}
    \omega _{0}\frac{e^{-\gamma _{+}\left|t\right|}+e^{-\gamma _{-}\left|t\right|}}{2}&=(\zeta /2)\phi _{p}(t)+\omega _{0}\phi _{q}(t)\\
    \intertext{and}
    \omega _{0}^{2}\frac{e^{-\gamma _{+}\left|t\right|}-e^{-\gamma _{-}\left|t\right|}}{2i\omega _{1}}&=-\omega _{0}\phi _{p}(t),
  \end{align}
\end{subequations}
we obtain
\begin{subequations}
  \begin{align}
    \mathcal{S}(t)&=S_{p}\phi _{p}(t)+S_{q}\phi _{q}(t)+\sum _{k=1}^{K}S_{k}e^{-\nu _{k}\left|t\right|}\\
    \intertext{and}
    \mathcal{A}(t)&=A_{p}\phi _{p}(t).
  \end{align}
\end{subequations}
where
\begin{subequations}
  \begin{align}
    S_{p}&\equiv \frac{2\lambda \zeta }{\beta \hbar }\left(2\sum _{k=1}^{K}\frac{\eta _{k}\omega _{0}\nu _{k}^{2}}{(\omega _{0}^{2}+\nu _{k}^{2})^{2}-\zeta ^{2}\nu _{k}^{2}}\right),\\
    S_{q}&\equiv \frac{2\lambda }{\beta \hbar }\left(1+2\sum _{k=1}^{K}\frac{\eta _{k}\omega _{0}^{2}(\omega _{0}^{2}+\nu _{k}^{2})}{(\omega _{0}^{2} + \nu _{k}^{2})^{2}-\zeta ^{2}\nu _{k}^{2}}\right)\\
    S_{k}&\equiv 2\eta _{k}\frac{2\lambda }{\beta \hbar }\left(\frac{\nu _{k}\omega _{0}^{2}\zeta }{(\nu _{\kappa }^{2}+\omega _{0}^{2})^{2}-\zeta ^{2}\nu _{\kappa }^{2}}\right),\quad \quad (k=1,\dots ,K)\\
    \intertext{and}
    A_{p}&\equiv \lambda \omega _{0}.
  \end{align}
\end{subequations}

The functions $\phi _{p}(t)$ and $\phi _{q}(t)$ satisfy
\begin{align}
  \partial _{t}
  \begin{pmatrix}
    \phi _{p}(t)\\
    \phi _{q}(t)\\
  \end{pmatrix}
  &=-\bm{\gamma }_{p,q}
  \begin{pmatrix}
    \phi _{p}(t)\\
    \phi _{q}(t)\\
  \end{pmatrix},&
  \bm{\gamma }_{p,q}&\equiv 
  \begin{pmatrix}
    \zeta &\omega _{0}\\
    -\omega _{0}&0\\
  \end{pmatrix},
\end{align}
and $\mathcal{S}(t)$ and $\mathcal{A}(t)$ can be rewritten in the form of Eqs.~\eqref{eq:S-expansion} and \eqref{eq:A-expansion} as
\begin{subequations}
  \begin{align}
    \mathcal{S}(t)&=
    \begin{pmatrix}
      0 & 1\\
    \end{pmatrix}
    \bm{s}_{p,q}
    \begin{pmatrix}
      \phi _{p}(t) \\
      \phi _{q}(t) \\
    \end{pmatrix}
    +\sum _{k=1}^{K}S_{k}e^{-\nu _{k}\left|t\right|}
    \label{eq:S-brownian-2x2}\\
    \intertext{and}
    \mathcal{A}(t)&=
    \begin{pmatrix}
      0 & 1\\
    \end{pmatrix}
    \bm{a}_{p,q}
    \begin{pmatrix}
      \phi _{p}(t)\\
      \phi _{q}(t)\\
    \end{pmatrix}.
  \end{align}
\end{subequations}
Here, $2\times 2$ matrices $\bm{s}_{p,q}$ and $\bm{a}_{p,q}$ are defined as
\begin{subequations}
  \begin{align}
    \bm{s}_{p,q}&\equiv -\frac{S_{p}}{\omega _{0}}\cdot 
    \bm{\gamma }_{p,q}
    +S_{q}\cdot \bm{1}
    \intertext{and}
    \bm{a}_{p,q}&\equiv -\frac{A_{p}}{\omega _{0}}\cdot \bm{\gamma }_{p,q},
  \end{align}
\end{subequations}
and these clearly commute with $\bm{\gamma }_{p,q}$.
The second term in Eq.~\eqref{eq:S-brownian-2x2} leads a $K\times K$ diagonal block matrix.

\section{Coefficients of \texorpdfstring{$\mathcal{S}(t)$}{S(t)} and \texorpdfstring{$\mathcal{A}(t)$}{A(t)} in Sec.~\ref{sec:example3}}
\label{sec:decomposition-brownian2}
In the Fano spectral decomposition scheme, the Bose--Einstein distribution function is decomposed as \cite{cui2019jcp}
\begin{align}
  n_{\mathrm{BE}}(\omega )+\frac{1}{2}&\simeq \frac{1}{\beta \hbar \omega }+\sum _{j}\frac{b_{j}a_{j}\omega /T_{j}}{[1+(a_{j}\omega /T_{j})^{2}]^{m_{j}}}.
  \label{eq:n-decomposition-fano}
\end{align}
Here, we have unified the high--temperature/low--temperature parts in Ref.~\onlinecite{cui2019jcp}.
By inserting Eqs.~\eqref{eq:brownian-alpha} and \eqref{eq:n-decomposition-fano} into Eqs.~\eqref{eq:S} and \eqref{eq:A} and by using the residue theorem, we obtain correlation functions $\mathcal{S}(t)$ and $\mathcal{A}(t)$.
In $\mathcal{S}(t)$, non--exponential basis functions $\phi _{\gamma _{j},l}(t)=t^{l}\cdot e^{-\gamma _{j}t}$ ($1\leq l\leq m_{j}$) appear as results of higher--order poles in Eq.~\eqref{eq:n-decomposition-fano} with $m_{j}\geq 1$.
The basis functions satisfy time evolution equation
\begin{align}
  \partial _{t}
  \begin{pmatrix}
    \phi _{\gamma _{j},0}(t)\\
    \phi _{\gamma _{j},1}(t)\\
    \vdots \\
    \phi _{\gamma _{j},m_{j}}(t)\\
  \end{pmatrix}
  &=-
  \begin{pmatrix}
    \gamma _{j}&&&\bm{0}\\
    -1&\gamma _{j}\\
    &\ddots &\ddots \\
    \bm{0}&&-m_{j}&\gamma _{j}\\
  \end{pmatrix}
  \begin{pmatrix}
    \phi _{\gamma _{j},0}(t)\\
    \phi _{\gamma _{j},1}(t)\\
    \vdots \\
    \phi _{\gamma _{j},m_{j}}(t)\\
  \end{pmatrix}.
  \label{eq:polynomial-exponential-basis}
\end{align}.

\section{Construction of Volterra--type integro--differential equation Eq.~\eqref{eq:volterra}}
\label{sec:volterra}
By inserting the total Hamiltonian with the RWA,
\begin{widetext}
  \begin{align}
    H^{\mathrm{tot}}=\hbar \Omega _{e}\hat{a}^{+}\hat{a}^{-}-\sum _{\xi }\frac{g_{\xi }}{\sqrt {\mathstrut 2}}\left(\hat{a}^{-}\hat{b}_{\xi }^{+}+\hat{a}^{+}\hat{b}_{\xi }^{-}\right)+\sum _{\xi }\hbar \omega _{\xi }\hat{b}_{\xi }^{+}\hat{b}_{\xi }^{-}+E_{\mathrm{vac}},
  \end{align}  
\end{widetext}
and Eq.~\eqref{eq:total-wavefunction} into the Schr\"{o}dinger equation, we obtain infinite number of simultaneous equations
\begin{align}
  \left\{
  \begin{aligned}
    \partial _{t}c_{e,0}(t)&=-i\left(\Omega _{e}+E_{\mathrm{vac}}/\hbar \right)c_{e,0}(t)+i\sum _{\xi }\frac{g_{\xi }}{\sqrt {\mathstrut 2}\hbar }c_{g,\xi }(t)\\
    \partial _{t}c_{g,\xi }(t)&=-i\left(\omega _{\xi }+E_{\mathrm{vac}}/\hbar \right)c_{g,\xi }(t)+i\frac{g_{\xi }}{\sqrt {\mathstrut 2}\hbar }c_{e,0}(t).
  \end{aligned}
  \right.
\end{align}
By introducing interaction picture
\begin{align}
  \left\{
  \begin{aligned}
  \tilde{c}_{e,0}(t)&=e^{+i\left(\Omega _{0}+E_{\mathrm{vac}}/\hbar \right)(t-t_{0})}c_{e,0}(t)\\
  \tilde{c}_{g,\xi }(t)&=e^{+i\left(\omega _{\xi }+E_{\mathrm{vac}}/\hbar \right)(t-t_{0})}c_{g,\xi }(t),
  \end{aligned}
  \right.
\end{align}
The set of equations can be rewritten as
\begin{subequations}
  \begin{align}
    \partial _{t}\tilde{c}_{e,0}(t)&=i\sum _{\xi }\frac{g_{\xi }}{\sqrt {\mathstrut 2}\hbar }e^{-i\left(\omega _{\xi }-\Omega _{e}\right)(t-t_{0})}\tilde{c}_{g,\xi }(t)\\
    \intertext{and}
    \partial _{t}\tilde{c}_{g,\xi }(t)&=i\frac{g_{\xi }}{\sqrt {\mathstrut 2}\hbar }e^{+i\left(\omega _{\xi }-\Omega _{e}\right)(t-t_{0})}\tilde{c}_{e,0}(t).
    \label{eq:c-tilde-2}
  \end{align}
\end{subequations}
Because Eq.~\eqref{eq:c-tilde-2} is solved as
\begin{align}
  \tilde{c}_{g,\xi }(t)&=\int _{t_{0}}^{t}\!ds\,i\frac{g_{\xi }}{\sqrt {\mathstrut 2}\hbar }e^{+i\left(\omega _{\xi }-\Omega _{e}\right)(s-t_{0})}\tilde{c}_{e,0}(s),
\end{align}
the time evolution of $\tilde{c}_{g,0}$ is expressed as
\begin{align}
  \partial _{t}\tilde{c}_{e,0}(t)&=-\sum _{\xi }\frac{g_{\xi }^{2}}{2\hbar ^{2}}\int _{t_{0}}^{t}\!ds\,e^{-i\left(\omega _{\xi }-\Omega _{e}\right)(t-s)}\tilde{c}_{e,0}(s).
\end{align}
By using the spectral density $\mathcal{J}(\omega )$, this equation can be rewritten as Eq.~\eqref{eq:volterra}.

\let\emph=\textit
\bibliography{tikeda_arXiv}

\end{document}